\documentclass[pdflatex,sn-nature,iicol]{sn-jnl}

\usepackage{graphicx}%
\usepackage{multirow}%
\usepackage{amsmath,amssymb,amsfonts}%
\usepackage{amsthm}%
\usepackage{mathrsfs}%
\usepackage[title]{appendix}%
\usepackage{xcolor}%
\usepackage{textcomp}%
\usepackage{manyfoot}%
\usepackage{booktabs}%
\usepackage{algorithm}%
\usepackage{algorithmicx}%
\usepackage{algpseudocode}%
\usepackage{listings}%
\usepackage{eso-pic,xcolor}
\usepackage{comment}
\usepackage{arydshln}
\usepackage{braket}

\theoremstyle{thmstyleone}%
\theoremstyle{thmstyletwo}%

\theoremstyle{thmstylethree}%
\newcommand{\kB}{k_{\mathrm{B}}}
\newcommand{\Otwo}{\mathrm{O}_2}
\newcommand{\Hethree}{^3\mathrm{He}}
\newcommand{\Hefour}{^4\mathrm{He}}
\newcommand{\aTa}{\alpha\text{-Ta}}
\newcommand{\bTa}{\beta\text{-Ta}}
\newcommand{\mTa}{\alpha/\beta\text{-Ta}}
\newcommand{\CFfour}{\mathrm{CF}_4}
\newcommand{\omegar}{\omega_{\mathrm{r}}}
\newcommand{\omegaq}{\omega_{\mathrm{q}}}

\newcommand{\fr}{f_{\mathrm{r}}}
\newcommand{\fzero}{f_0}
\newcommand{\Strans}{S_{21}}
\newcommand{\pMS}{p_{\mathrm{MS}}}
\newcommand{\Pin}{P_{\mathrm{in}}}
\newcommand{\Qi}{Q_\mathrm{i}}
\newcommand{\Qc}{Q_\mathrm{c}}
\newcommand{\Ql}{Q_\mathrm{l}}
\newcommand{\Qe}{Q_\mathrm{e}}
\newcommand{\Qzero}{Q_0}
\newcommand{\Qother}{Q_{\mathrm{other}}}
\newcommand{\Qtlsz}{Q_\mathrm{TLS,0}}
\newcommand{\navg}{\langle n_{\mathrm{ph}} \rangle}
\newcommand{\nc}{n_\mathrm{c}}
\newcommand{\Tone}{T_1}
\newcommand{\TPurcell}{T_\mathrm{1,Purcell}}
\newcommand{\Techo}{T_2^{\mathrm{echo}}}
\newcommand{\Tc}{T_{\mathrm{c}}}
\newcommand{\Tact}{T_{\mathrm{act}}}
\newcommand{\tandeff}{\tan\delta_\mathrm{eff}}

\newcommand{\degC}{^\circ\mathrm{C}}
\newcommand{\us}{\mu\mathrm{s}}
\newcommand{\ns}{\mathrm{ns}}
\newcommand{\GHz}{\mathrm{GHz}}
\newcommand{\MHz}{\mathrm{MHz}}
\newcommand{\dB}{\mathrm{dB}}

\newcommand{\mK}{\mathrm{mK}}
\newcommand{\Kelvin}{\mathrm{K}}
\newcommand{\mBar}{\mathrm{mBar}}
\newcommand{\muBar}{\mu\mathrm{Bar}}
\newcommand{\micron}{\mu\mathrm{m}}
\newcommand{\mm}{\mathrm{mm}}
\newcommand{\nm}{\mathrm{nm}}
\newcommand{\sccm}{\mathrm{sccm}}
\newcommand{\kV}{\mathrm{kV}}
\newcommand{\W}{\mathrm{W}}

\begin{document}

\title[Article Title]{Robust growth of {\unboldmath $\alpha$}-tantalum on sapphire at room temperature using atomically thin titanium seed layer}

\author{\fnm{Anand} \sur{Kamlapure}\textsuperscript{1,2,*}}
\author{\fnm{Lucas} \sur{Hak}\textsuperscript{1,2,*}}
\author{\fnm{Kishore Kumar} \sur{Thiyagarajan}\textsuperscript{1,2}}
\author{\fnm{Hozanna} \sur{Miro}\textsuperscript{2}}
\author{\fnm{Marc} \sur{Beekman}\textsuperscript{3}}
\author{\fnm{Matvey} \sur{Finkel}\textsuperscript{1,2}}
\author{\fnm{Leonardo} \sur{DiCarlo}\textsuperscript{1,2,\textdagger}}

\affil[1]{\orgname{QuTech, Delft University of Technology}, \orgaddress{\street{P.O. Box 5046}, \postcode{2600 GA}, \city{Delft}, \country{The Netherlands}}}

\affil[2]{\orgname{Kavli Institute of Nanoscience, Delft University of Technology}, \orgaddress{\street{P.O. Box 5046}, \postcode{2600 GA}, \city{Delft}, \country{The Netherlands}}}

\affil[3]{\orgname{Netherlands Organisation for Applied Scientific Research (TNO)}, \orgaddress{\street{P.O. Box 96864}, \postcode{2509 JG}, \city{The Hague}, \country{The Netherlands}}}


\abstract{

Tantalum (Ta) based superconducting circuits have attracted attention for their low microwave loss. However, reproducible growth of high-quality {\unboldmath $\alpha$}-phase Ta films remains challenging because film properties are highly sensitive to growth conditions. We present a robust, room-temperature route for {\unboldmath $\aTa$} growth on sapphire substrates using an ultrathin Ti seed layer, which significantly widens the accessible deposition parameter space. We compare this approach to Ta films grown at high temperature without a seed layer and at room temperature with a Nb seed. We assess films using structural and surface characterization and quantify microwave loss with coplanar-waveguide (CPW) resonators measured in the quantum regime. We obtain two-level-system (TLS)-limited quality factors comparable to the state of the art for Ta-on-sapphire CPW resonators with similar geometries. In addition, temperature-dependent measurements reveal that high-temperature-grown films and films subjected to extended buffered-oxide etch treatment exhibit additional temperature activated loss, highlighting the role of growth route and post-processing in microwave performance. Our results establish ultrathin seed-layer engineering as a reliable alternative route to reproducible, low-loss {\unboldmath $\aTa$} films and motivate closer study of the interfacial mechanisms that govern Ta phase formation.
}

\maketitle

\begingroup
\renewcommand{\thefootnote}{}
\footnotetext{\textsuperscript{\textdagger}Email: l.dicarlo@tudelft.nl}
\footnotetext{\textsuperscript{*}These authors contributed equally to this work.}
\endgroup

\section{Introduction}\label{sec1}

Superconducting quantum circuits promise long coherence, fast gate operations, and scalability, making them a leading platform for quantum information processing ~\cite{DeLeon21,Rosenberg17,Place21,Bal24}. Tantalum (Ta) has emerged as a particularly promising material for these devices, with advances in materials optimization and device engineering enabling coherence times approaching the millisecond range~\cite{Bland25,Ganjam24}. This is attributed to the low microwave loss associated with its stable native oxide ~\cite{Bland25,Ganjam24,Place21,Wang22,Bal24,Lozano25,Marcaud25,McLellan23}.
Despite this progress, reproducible growth of high-quality $\alpha$-phase Ta remains challenging because its structural and superconducting properties are highly sensitive to growth conditions~\cite{Alegria25,Bu25,Karuppannan25,VanSchijndel25,Wu23,Zhou23,Jia23}. In practice, it is comparatively easy to obtain films with mixed $\alpha/ \beta$ phase content, whereas realization of phase-pure $\aTa$ requires tighter control over substrate, temperature, interface quality, and growth kinetics~\cite{Dhundhwal25,Bu25}. Although recent studies show that $\beta$-containing Ta films can support high resonator and qubit quality factors~\cite{Dhundhwal25, Joshi26}, phase-pure $\aTa$ remains desirable from the standpoint of process consistency and materials optimization, especially because the relationship between phase composition, disorder, and microwave loss is not fully resolved~\cite{Bahrami26, Dhundhwal25}. Recent studies have explored a broad process space, including direct growth on heated substrates, cryogenic deposition, different seed layers, different substrate materials, and alternative sputter gases. For example, direct growth on heated Si or on selected seed layers such as Nb can stabilize $\aTa$, while the use of Kr instead of Ar as a sputter gas enables reduced substrate temperature required for $\aTa$ formation~\cite{Urade24,Marcaud25,Singer24,Karuppannan25,VanSchijndel25,Olszewski26}.

At the same time, achieving nominally phase-pure $\aTa$ does not by itself ensure optimal microwave performance ~\cite{Dhundhwal25, Alegria23, McLellan23, Oh24, Murthy25,Moshel25}. Recent reports show that Ta remains sensitive to subtle growth and process induced disorder. Surface chemistry can introduce additional loss even in otherwise phase-pure films; in particular, surface hydrogenation during HF-based oxide removal step has been identified as a source of increased microwave loss, with performance recovering after vacuum annealing~\cite{Lozano25}. Further, thermally activated microwave loss in Ta resonators has been shown to arise from vortex motion, appearing in clean-limit films and being suppressed in dirtier films where structural defects provide stronger vortex pinning~\cite{Bahrami26}.
Together, these findings indicate that improvements in crystallinity and purity do not necessarily eliminate all relevant loss channels.
More broadly, these results show that the relationship between phase purity, disorder, and microwave loss in Ta remains incompletely understood, underscoring the need to connect materials growth with resonator performance, which serves as a practical proxy for qubit coherence in quantum processors because both are limited by similar material-related loss channels~\cite{McRae20,Crowley23}.

We address this challenge through ultrathin Ti seed-layer engineering. We define one monolayer (1~ML) of Ti as a nominal coverage-equivalent thickness of $2.34~\text{\AA}$, corresponding to one close-packed $\mathrm{Ti}(0001)$ atomic layer on $\mathrm{Al_2O_3}(0001)$~\cite{Sondergaard04}. The ML values used below therefore denote calibrated nominal thicknesses. We show that a ML-scale Ti seed layer enables reliable room-temperature growth of $\aTa$ over a broad range of deposition parameters. In contrast to approaches that are sensitive to growth parameters to stabilize $\aTa$ and suppress $\bTa$ formation~\cite{Majer24,Zikiy26}, our method reproducibly yields $\aTa$ under different growth conditions and in different deposition environments, all at room temperature. We benchmark this approach against high-temperature-grown films without a seed layer and include a representative Nb-seeded room-temperature film as an additional comparison. We evaluate film quality and microwave loss using structural, surface, and coplanar-waveguide (CPW) resonator measurements. We further examine the temperature-dependent microwave response of the resonators from the low-temperature regime dominated by two-level-system (TLS) loss into the quasiparticle (QP) dominated regime, revealing an additional thermally activated loss channel in a subset of devices. Beyond establishing an alternative practical route to robust room-temperature $\aTa$
growth, our study also helps clarify how purity and disorder-related material characteristics are reflected in resonator performance. Finally, we assess the compatibility of the optimized Ti-seeded $\aTa(110)$ film by characterizing a test device with 10 transmons.

\section{Ultrathin Ti seed layer to stabilize $\aTa$ at room temperature}
\begin{figure*}[htbp]
\centering
\includegraphics[width=\linewidth]{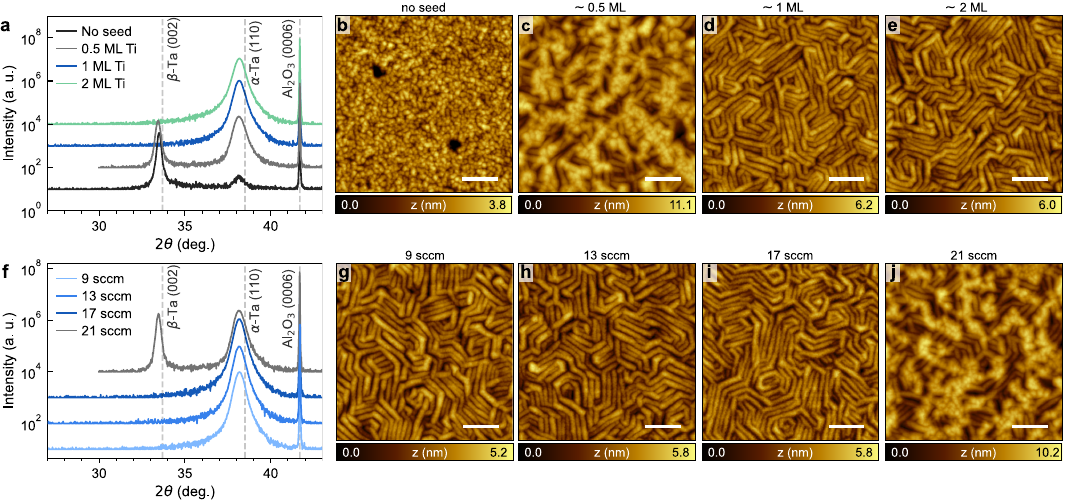}
\caption{\textbf{Ti seed layer stabilizes phase-pure $\aTa$ growth.}
XRD and AFM images showing the evolution of Ta film phase and morphology as a function of Ti seed-layer thickness and Ar flow. (a–e) At a fixed Ar flow of $17~\sccm$, without a seed layer, Ta films form predominantly $\beta$-phase with traces of $\alpha$-phase. Introducing $\sim0.5~\mathrm{ML}$ Ti seed layer leads to mixed-phase growth, while increasing the thickness to 1~ML and 2~ML yields phase-pure $\aTa$ with a pronounced (110) texture. Corresponding AFM images show a transition from irregular grain morphology to tightly packed, hexagonally ordered elongated grains. (f–j) Films grown with a 1~ML Ti seed layer remain phase-pure $\aTa(110)$ over a broad range of Ar flow ($9–17~\sccm$), with deviation observed at $21~\sccm$. Corresponding AFM images show largely unchanged morphology across the stable growth window and phase-separated regions at high flow. Scale bars in all AFM images are 100~nm.
}
\label{fig:fig1}
\end{figure*}
We first investigate the role of the Ti seed layer in the resulting phase of sputtered Ta films grown at room temperature. Thin-film deposition details are provided in the Methods. Figure~\ref{fig:fig1} summarizes the film evolution with seed-layer thickness and sputtering conditions using X-ray diffraction (XRD) and atomic force microscopy (AFM). Without a seed layer, the deposited Ta films form predominantly $\beta$ phase with traces of $\alpha$ phase. Introducing $\sim$0.5~ML Ti seed layer leads to mixed-phase growth, whereas $\sim$1~ML or $\sim$2~ML yields phase-pure $\aTa$ with a pronounced (110) grain texture (Fig.~\ref{fig:fig1}d, e). This progression shows that a monolayer-scale Ti seed layer is sufficient to drive the growth from predominantly $\beta$-phase to single-phase $\aTa$. The corresponding $\aTa(110)$ reflection appears at $2\theta={38.2}^\circ$, slightly below the literature value, which we attribute to in-plane compressive stress~\cite{Jia23}. The AFM images show a corresponding evolution in surface morphology. Films grown without Ti, as well as those with sub-monolayer Ti, retain the irregular grain structure associated with $\beta$-containing growth, including elongated-grain regions separated by circular grains. By contrast, films grown with 1~ML and 2~ML Ti exhibit tightly packed, hexagonally ordered elongated grains, consistent with the formation of $\aTa(110)$.

Having established that 1~ML Ti is sufficient to stabilize $\aTa$, we next examine the robustness of this phase against variations in Ar flow. As shown in Fig.~\ref{fig:fig1}g--j, films grown with a 1~ML Ti seed layer remain phase-pure $\aTa(110)$, with largely unchanged characteristic morphology from $9$ to $17~\sccm$. Deviations from phase purity, evidenced by phase-separated regions, appear by $21~\sccm$. This indicates that once the Ti layer reaches approximately one monolayer, the growth becomes considerably less sensitive to deposition conditions than in films grown without Ti. Previous first-principles studies have examined the delicate balance of interfacial energetics involved in Ta phase selection~\cite{Bu25}. Our results suggest that even a single layer of Ti may reduce the interfacial stress and promote $\aTa(110)$, although clarifying the microscopic mechanism would require first-principles analysis beyond the scope of this work.

\begin{figure*}[t]
\centering
\includegraphics[width=\linewidth]{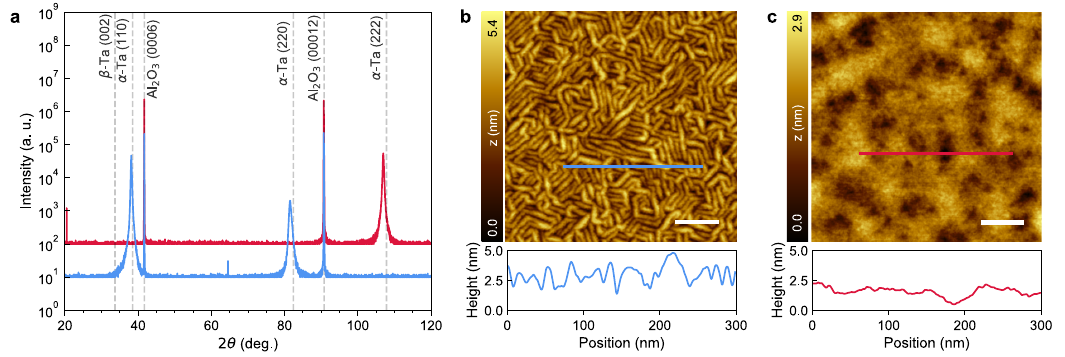}
\caption{\textbf{Comparison of room-temperature and high-temperature $\aTa$ growth.}
XRD and AFM characterization of $\aTa$ films grown using different approaches. (a) Wide-range XRD spectra of (blue)  room-temperature-grown Ti-seeded $\aTa$ and (red) high-temperature-grown $\aTa$. The Ti-seeded film is strongly (110)-textured, whereas the high-temperature-grown film is (111)-textured. The (111) orientation is identified by the (222) peak~\cite{Cullity01} because the (111) reflection is forbidden for $\aTa$. (b,c) Corresponding AFM images show distinct surface morphologies. (b) The $\aTa(110)$ films exhibit the typical surface corrugation of the $\aTa(110)$ phase, with elongated grain structure and higher roughness ($\sim0.73~\nm$). The $\aTa(111)$ films are smoother ($\sim0.39~\nm$) but display a high density of surface defects, seen as darker features. Line profiles below AFM images highlight the contrast in surface morphology between the two growth modes. Scale bars in (b) and (c) are $100~\nm$.}
\label{fig:fig2}
\end{figure*}

To further benchmark the room-temperature Ti-seeded films, we deposit $\aTa$ in a second sputtering system (see Methods) using substantially different growth conditions and compare the resulting films with $\aTa$ grown using reported high-temperature recipes (Table \ref{tab:growth_parameters}). As shown in Fig.~\ref{fig:fig2}, the Ti-seeded room-temperature films exhibit strongly textured $\aTa(110)$, whereas the high-temperature-grown films show phase-pure $\aTa$ with dominant (111) orientation. These distinct textures are accompanied by considerably different surface morphologies. The $\aTa(110)$ films have a root-mean-square (RMS) roughness of $\sim0.73~\nm$ and exhibit elongated grain structure, while the high-temperature-grown $\aTa(111)$ films have $\sim0.39~\nm$ and display visible surface defects. The contrast is further evident in the line profiles. Although these defects cannot be resolved in detail here, their appearance is consistent with previous reports associating such features with screw dislocations in high-temperature-grown $\aTa$ films~\cite{Alegria25}.

Together, these results show that a monolayer-scale Ti seed layer enables reliable room-temperature growth of phase-pure $\aTa(110)$. This effect is, however, restricted to the ultrathin limit. In the second system, increasing the Ti thickness to $\sim$1.4~ML ($3.3 \pm 0.3$~\AA) already results in mixed-phase growth, whereas in the first system phase-pure $\aTa$ persists up to $\sim$2~ML ($5.6 \pm 0.5$~\AA). The upper bound of the Ti thickness window therefore depends on the deposition system, presumably reflecting differences in the seed-layer growth conditions. While the full parameter space is not exhaustively explored here, the robustness of this approach is evident from its reproducibility across two sputtering systems and over the range of deposition conditions investigated~(Supplementary Section~\ref{supsec:depo_details}).

\section{Low temperature resonator characterization}

\begin{figure*}[t]
\centering
\includegraphics[width=\linewidth]{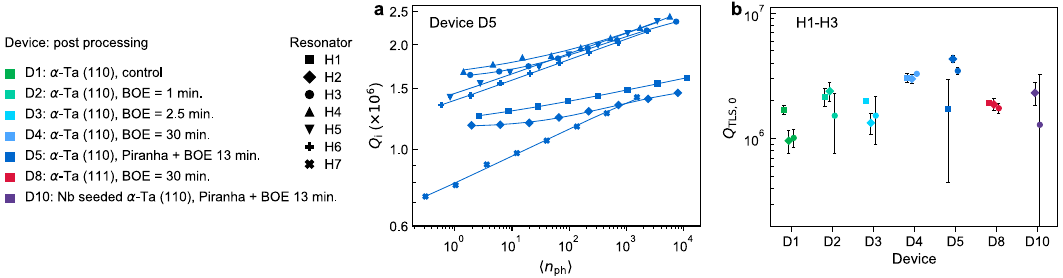}
\caption{\textbf{Low-temperature TLS-limited performance of $\aTa$ films.}
Low-temperature characterization of CPW resonators fabricated on Ta films. The legend identifies the film type and post-processing condition associated with each device. Unless labeled Nb-seeded, $\aTa(110)$ films are Ti-seeded, and $\aTa(111)$ films are grown at high temperature without a seed layer. (a) Power-dependent internal quality factor $\Qi$ for resonators using the optimized Ti-seeded $\aTa(110)$ film, together with a fit using the TLS saturation model. (b) Extracted $\Qtlsz$ for the widest resonators (H1--H3, $20~\micron$ gap width) across different films and cleaning conditions, showing similar TLS-limited performance. The highest $\Qtlsz$ is observed in resonators using the optimized Ti-seeded $\aTa(110)$ film. Large uncertainties in $\Qtlsz$ for some resonators reflect a nearly power-independent $\Qi$, for which the TLS model fit is poorly constrained (see Supplementary Section~\ref{supsec:pow_dep_tls}).}
\label{fig:fig3}
\end{figure*}

We first benchmark the microwave performance of the Ta films by measuring CPW resonators in a dilution refrigerator at base temperature ($\sim20~\mK$) and analyzing their power-dependent internal quality factor $\Qi$ to extract the TLS-limited quality factor, $\Qtlsz$~\cite{Crowley23, Moshel25, Lozano25}. Details of the power-dependent fitting model are given in Methods. To vary the dielectric participation ratio (PR), we fabricate resonator geometries with different CPW center-conductor and gap widths (Supplementary Section~\ref{supsec:resonator_layout}), with corresponding PR calculations provided in the Supplementary Section~\ref{supsec:PR}. We use the metal--substrate PR, $\pMS$, as a representative ratio, since the dominant interface participations in CPW geometries are expected to scale together with geometry ~\cite{Wenner11,Wang15,Murray18,Martinis22,VanDamme24}. To examine film and processing dependent effects alongside this geometrical variation, we compare $\aTa$ films grown at room temperature using Ti and Nb seed layers, and ones grown at high temperature. These devices were subjected to a range of post-fabrication cleaning conditions, including buffered oxide etch (BOE)-based treatments, as summarized in Table~\ref{tab:device_cleaning_conditions}. Figure~\ref{fig:fig3}a shows one such power sweep for the optimized-cleaning, room-temperature Ti-seeded $\aTa(110)$ film, together with a fit using the TLS saturation model. Figure~\ref{fig:fig3}b summarizes the extracted $\Qtlsz$ for the widest resonators H1--H3 ($20~\micron$ gap) across eight devices spanning different film types and post-fabrication cleaning methods. The complete set of power-dependent data, corresponding TLS fits, and extracted parameters for all resonators and devices is provided in Supplementary Section~\ref{supsec:pow_dep_tls}.

For the optimized room-temperature-grown, Ti-seeded $\aTa(110)$ film, $\Qtlsz >3.0\times10^6$ for the widest resonators. Across the measured films and cleaning conditions, the Ti-seeded $\aTa(110)$ films show low-power TLS-limited performance comparable to that of the high-temperature-grown $\aTa(111)$ reference film. These low-temperature measurements therefore establish that an ultrathin Ti seed layer provides a reliable room-temperature route to $\aTa$ films with microwave performance comparable to established growth methods~\cite{Moshel25,Alegria25,Crowley23,Bu25,Dhundhwal25, Bahrami26}. We next turn to the temperature dependence of resonator performance, where differences between growth routes and surface treatments become more apparent.

\section{Temperature evolution of $\Qi$}
\label{sec:tdep}
\begin{figure*}[htbp]
\centering
\includegraphics[width=\linewidth]{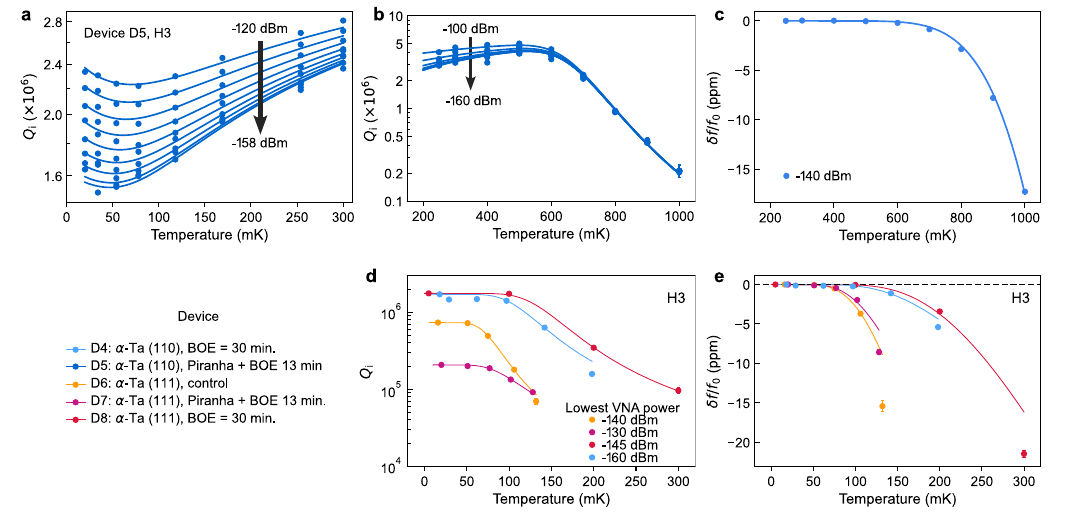}
\caption{\textbf{Temperature-dependent loss and frequency shifts.}
Temperature dependence of internal quality factor $\Qi$ and fractional frequency shift  $\delta f/\fzero$. All measurements shown correspond to resonator H3. (a) Extracted $\Qi$ in device D5 (optimized Ti-seeded $\aTa(110)$ film) as a function of temperature up to $300~\mK$ in a dilution refrigerator, at varying incident power. In this low temperature range, the full dataset is well described by a TLS model. (b) When extending the temperature range to $1~\Kelvin$ using a $\Hethree$ refrigerator, $\Qi$ is best described by a combined TLS and quasiparticle model. (c) Corresponding  $\delta f/\fzero$, and fit of the quasiparticle model. (d,e) Temperature dependence of (d) low-power $\Qi$ and (e) $\delta f/\fzero$ for resonator devices showing thermally-acivated additional loss. All resonators using high-temperature-grown $\aTa(111)$ films show activated loss regardless of cleaning condition, as does the resonator using room-temperature-grown, Ti-seeded $\aTa(110)$ film subjected to 30~min BOE cleaning.}
\label{fig:fig4}
\end{figure*}

We next examine the temperature dependence of the resonator $\Qi$ to probe additional loss mechanisms beyond the low-temperature TLS regime. Figure~\ref{fig:fig4}a shows the temperature dependence of $\Qi$ for the optimized room-temperature-grown Ti-seeded $\aTa(110)$ film, measured in a dilution refrigerator from base temperature up to $\sim300~\mK$. The data are well described by the temperature and power dependent TLS model with empirical exponents $\beta_1$ and $\beta_2$, without invoking quasiparticle loss~\cite{Crowley23} (see Methods). Extended temperature and power-dependent datasets for all samples, including the representative Nb-seeded film, are shown in Supplementary Fig.~\ref{fig:Qi_pow_temp}, with the corresponding best-fit parameter values summarized in Supplementary Fig.~\ref{fig:beta1_beta2_model}. The extracted $\Qtlsz$ is comparable to that obtained from power-dependent measurements at base temperature, with the latter yielding slightly higher values (see Supplementary Section~\ref{supsec:Qtls0_comparison}).

To explore whether additional temperature-dependent loss channels emerge above the low-temperature TLS regime, we extend the measurement range using a $\Hethree$ cryostat. Measurements were performed on the same optimized Ti-seeded $\aTa(110)$ film from $\sim250~\mK$ to $1~\Kelvin$, as shown in Fig.~\ref{fig:fig4}b. In this regime, the temperature dependence of $\Qi$ can be described by including both TLS contributions and thermally generated quasiparticles~\cite{Turneaure91,Zmuidzinas12,Anlage21}. Because quasiparticle loss dominates and TLS saturation is thermally suppressed at these temperatures, $\beta_1$ is not constrained by the data and is therefore fixed to that extracted from fits to the dilution-refrigerator data, while $\Tc$ is fixed to its measured value (see Methods and Supplementary Section~\ref{supsec:110_vespa}). The corresponding frequency shift, $\delta f/f$, shown in Fig.~\ref{fig:fig4}c, is also well captured by the quasiparticle model. The extended temperature-dependent measurements and corresponding fits are shown in Supplementary Fig.~\ref{fig:tdep_vespa}, with the extracted fit parameters summarized in Supplementary Fig.~\ref{fig:tdep_vespa_params}. The resulting $\Qtlsz$ is comparable to that obtained from the dilution-refrigerator measurements (see Supplementary Section~\ref{supsec:Qtls0_comparison}). Together, these results indicate that the optimized Ti-seeded $\aTa(110)$ film follows the expected temperature evolution over the full measured range.

This conventional TLS-quasiparticle description breaks down for a different subset of devices. Additional temperature-dependent loss is observed in the high-temperature-grown $\aTa(111)$ films and in the Ti-seeded $\aTa(110)$ film subjected to prolonged buffered-oxide treatment. As shown in Fig.~\ref{fig:fig4}d-e, the measured high-temperature-grown $\aTa(111)$ films, including the control device, the optimized-cleaning device, and the 30 min BOE-treated device, exhibit a pronounced reduction of low-power $\Qi$ above $\sim100~\mK$. A similar degradation is observed in Ti-seeded $\aTa(110)$ after prolonged 30 min buffered-oxide etching. However, among the room-temperature $\aTa(110)$ films, activated loss is not observed for the Ti-seeded films treated with moderate BOE or optimized cleaning. The representative Nb-seeded comparison film also shows no activated behavior, as seen in Supplementary Fig.~\ref{fig:Qi_pow_temp}l-m. The observed decrease cannot be captured by the TLS and quasiparticle models used above. The corresponding frequency shift also appears at temperatures well below the onset expected for the standard quasiparticle response (see Supplementary Section~\ref{supsec:110_vespa}) and instead follows an activated behavior. Extended data for the remaining resonators, together with a summary of the extracted activation temperatures, are shown in Supplementary Fig.~\ref{fig:tdep_dissipation}. We therefore describe the excess loss phenomenologically using an activated form, following the framework used for thermally activated loss in Ta resonators~\cite{Bahrami26}.

\section{Discussion}
The results presented establish two main findings. First, an ultrathin Ti seed layer enables a reliable room-temperature growth route to phase-pure $\aTa$ with microwave performance comparable to that obtained from high-temperature and Nb-seeded processes. Second, temperature-dependent measurements reveal an additional thermally activated loss channel in some devices, not captured by conventional TLS and quasiparticle models.

At low temperature, the optimized Ti-seeded $\aTa(110)$ films show the TLS-limited behavior, with $Q_\mathrm{TLS, 0}$ comparable to recent high-performing $\aTa$ resonators on sapphire substrate~\cite{Crowley23,Dhundhwal25,Moshel25}. Upon increasing temperature, the same films are well described by a combination of TLS and quasiparticle contributions, and the corresponding frequency shift follows the standard quasiparticle response~\cite{Crowley23,Dhundhwal25}. The agreement in both dissipative and reactive channels indicates that the dominant loss mechanisms in these films are captured by established models. The representative Nb-seeded $\aTa(110)$ sample shows broadly comparable TLS-limited behaviour, although a single sample does not support a systematic comparison between seed materials. Native niobium oxides, particularly Nb$_2$O$_5$, are known to host TLSs and contribute microwave loss in superconducting resonators and qubits~\cite{Verjauw21,Bal24}. The Ti-seeded approach therefore avoids the use of Nb as a seed material and its associated Nb-containing interfaces, while retaining reliable room-temperature $\aTa$ growth across a widened deposition parameter space.

As a device-level validation, we fabricated a ten-transmon device using the optimized Ti-seeded $\aTa(110)$ film corresponding to device D5 in Fig.~\ref{fig:fig3}. Standard time-domain characterization showed a device-averaged transmon relaxation time  $\Tone = 96\pm19~\us$, and highest single-transmon time-averaged $\Tone=121~\pm23~\us$. These measurements demonstrate that the room-temperature-grown Ti-seeded films are compatible with Josephson-junction processing and can support transmon relaxation times at the $100~\us$ level. Fabrication and measurement details are provided in Supplementary Section~\ref{supsec:century}, with the coherence measurements shown in Supplementary Fig.~\ref{fig:century}.

Returning to resonators, a qualitatively different behavior emerges in both high-temperature-grown $\aTa(111)$ films and in Ti-seeded $\aTa(110)$ films subjected to extended buffered-oxide treatment. For these samples, $\Qi$ degrades above $\sim150~\mK$ in a manner that cannot be described by TLS or quasiparticle models. The accompanying frequency shift follows an activated temperature dependence with the same activation scale as the dissipative response. Together, these observations identify an additional activated response while leaving its microscopic origin unresolved.

One possible framework for interpreting these observations is vortex-motion-induced loss. Recent work on Ta resonators has shown that thermally activated vortex motion can generate additional microwave dissipation, with a strong dependence on film cleanliness and disorder~\cite{Bahrami26}. In that picture, cleaner films exhibit enhanced temperature-dependent loss, consistent with weaker pinning and greater vortex mobility. The activated behavior observed here is consistent with a vortex-creep picture, in which vortices thermally depin from local pinning sites and dissipate energy under microwave drive. Such thermally activated flux motion has long been discussed in superconducting thin films and is commonly associated with an activation barrier and Arrhenius-type temperature dependence~\cite{Bahrami26,Coffey91,Vinokur90,Blatter94,Song09,Silva17}. However, confirming a vortex origin would require controlled magnetic-field-dependent measurements and a more detailed analysis of vortex trapping and pinning, which are beyond the scope of this study.

The fact that both high-temperature-grown $\aTa(111)$ films and Ti-seeded $\aTa(110)$ films subjected to extended buffered-oxide treatment exhibit similar degradation suggests that the relevant mechanism is not determined by crystallographic phase alone, but may also be influenced by the growth, post-processing steps, and defect and pinning landscape~\cite{Bahrami26,Alegria25,Lozano25,Murthy25}. At the same time, the observation of activated loss in the control $\aTa(111)$ device indicates that this behavior cannot be assigned solely to extended buffered-oxide exposure. Surface treatment nevertheless remains relevant. Hydrogen incorporation during oxide removal has been reported to introduce additional microwave loss in $\aTa$, with partial recovery after vacuum annealing~\cite{Lozano25}. In our case, extended buffered-oxide treatment is associated with degraded $\Qi$ in Ti-seeded $\aTa(110)$, indicating that post-processing can influence the loss landscape. However, the present measurements do not identify the chemical or structural origin of this change. Whether extended buffered-oxide treatment modifies the pinning landscape, involves hydrogen related changes, or acts through another process-induced mechanism remains unresolved.

These observations suggest that both growth conditions and post-processing play an important role in the elevated-temperature microwave response of Ta films. The interplay between film purity, defect structure, and possible vortex dynamics may therefore lead to competing effects, such that improvements in crystallinity and surface quality do not necessarily eliminate all relevant loss channels.

In conclusion, we have shown that an ultrathin Ti seed layer enables robust room-temperature growth of phase-pure $\aTa$ over a broad range of deposition conditions. CPW resonators fabricated from these films exhibit low-temperature microwave performance comparable to state-of-the-art $\aTa$ CPW resonators of similar geometry, including films grown at high temperature and Nb-seeded films grown at room temperature~\cite{Moshel25,Alegria25,Crowley23,Bu25,Bahrami26,Marcaud25,Li25b}, as well as to the representative room-temperature Nb-seeded $\aTa$ film included in this study. At the same time, temperature-dependent measurements reveal an additional thermally activated dissipation channel in high-temperature-grown films and after extended buffered-oxide treatment, which is absent in the cleaning optimized Ti-seeded films, showing that elevated-temperature microwave performance is sensitive to growth route and post-processing conditions. These results establish ultrathin Ti seed-layer engineering as a practical alternative route to reproducible, low-loss $\aTa$ films, while also motivating further study of activated loss mechanisms in Ta-based quantum circuits. From a materials perspective, the stabilization of phase-pure $\aTa$ by a monolayer-thick Ti seed suggests that phase selection of Ta on other substrates may also be controlled through atomic-scale interface engineering.

\section{Methods }
\subsection{Substrate preparation and thin film deposition}
Sapphire substrates were cleaned in a 2:1 piranha solution of $\mathrm{H_2SO_4}:\mathrm{H_2O_2}$ at $80\degC$ for approximately 20~min and then rinsed in multiple baths of deionized water. Selected substrates were additionally annealed in a $\mathrm{N}_2:\mathrm{O}_2$ atmosphere (4:1) at $1050\degC$ for about 2.5~h, followed by the same piranha cleaning procedure before film growth. No clear difference was observed between annealed and unannealed substrates in the stabilization of the $\alpha$ phase.

Ta thin films were deposited by magnetron sputtering in two systems, an Alliance Concept AC450 and an AJA International ATC 1800-V, to confirm reproducibility across deposition platforms. The two systems differ in source geometry, with the AC450 using a 4-inch sputter target in a planar cathode configuration and the AJA system using a 2-inch sputter target in a confocal cathode configuration. For Ti-seeded growth, the seed layer and Ta were deposited in the same chamber and without breaking vacuum. The Ti seed layer was deposited using both DC and RF sputtering, with consistent stabilization of $\aTa$. Its deposition rate was calibrated from AFM step-height measurements using a lift-off-defined thickness step, enabling controlled monolayer-scale growth. Ta was subsequently deposited by DC magnetron sputtering in Ar. All Ti-seeded Ta films discussed were grown at room temperature unless stated otherwise.

For comparison, high-temperature $\aTa$ films were deposited in the AJA system on heated substrates. The substrate was heated to $750\degC$ and held for $\sim$1~h before deposition to allow the chamber to reach a lower base pressure. Ta was then deposited directly onto the heated substrate by DC sputtering. Details of the deposition parameters, including gas flow, pressure, power, and system-specific configurations, are summarized in Supplementary Section~\ref{supsec:depo_details}.

\subsection{Sample characterization}
The deposited films were characterized by XRD and AFM. XRD measurements were performed on a Bruker D8 Discover system using Cu K$\alpha$ radiation ($\lambda = 1.5406$~\AA) in a coupled $\theta$--$2\theta$ configuration~\cite{Cullity01}. Diffraction patterns were acquired over a $2\theta$ range spanning $20^\circ$ to $120^\circ$,  with a $0.015^\circ$ step size, and the X-ray source was operated at $50~\kV$. Surface morphology was characterized by AFM using a Bruker Dimension FastScan system. Images were acquired in tapping mode under ambient conditions. The data were plane-corrected prior to analysis, and surface roughness was evaluated from the height channel.

\subsection{Device fabrication}
CPW resonators were subsequently fabricated for low-temperature microwave measurements. After Ta film growth, the wafer was coated with S1813 photoresist to protect the film during dicing. Prior to lithography, this resist was removed by sequential immersion in acetone and isopropanol (IPA). Electron-beam lithography was performed using AR-P 6200.13 resist to define the resonator patterns. After exposure, the resist was developed in pentyl acetate, followed by a short dip in xylene and rinsing in two beakers of IPA. The resist was then hard baked at $130\degC$ prior to etching.

Pattern transfer was carried out in a Sentech Etchlab 200 etcher using $\CFfour$ at $20~\sccm$ and $50~\W$ RF power. Before etching, a gentle $\Otwo$ plasma clean was applied to remove residual resist from the exposed Ta surface. After etching, a second gentle $\Otwo$ plasma clean was applied to remove fluorocarbon residues from the $\CFfour$ etch, and the chips were then coated with photoresist prior to a second dicing step to obtain the final device dimensions. The protective resist was then stripped using PRS-3000, followed by acetone and IPA.

To examine the effect of post-processing surface treatments, several devices were prepared. Control samples received no further wet cleaning after resist stripping, and were directly mounted on a printed circuit board (PCB) and wirebonded (Al with $1\%$ Si, $25~\micron$ diameter). A second set of devices was treated in 7:1 buffered oxide etch (BOE) for different durations, as discussed in the main text. In the BOE-only series, different etch times were used to assess the effect of oxide removal. A third set of devices was treated using the optimized cleaning procedure, consisting of a 20~min piranha clean with a 2:1 $\mathrm{H_2SO_4:H_2O_2}$ mixture followed by a 13~min dip in 7:1 BOE. This procedure gave the best low-temperature microwave performance for Ti-seeded $\aTa(110)$ films, and thus we refer to it as the optimized cleaning procedure throughout.

\subsection{Extraction of $\Qtlsz$}
\label{method:qtls}
To extract $\Qtlsz$, we fit a standard TLS saturation model~\cite{Ganjam24,Crowley23} to the  measured internal loss $1/\Qi$ as function of average photon number $\navg$ at base temperature,
\begin{equation}\label{eq:TLSSaturation}
\frac{1}{\Qi}=\frac{1}{\Qtlsz}
\frac{\tanh\!\left(\frac{\hbar\omegar}{2\kB T}\right)}
{\sqrt{1+\left(\frac{\navg}{\nc}\right)^{\beta}}}
+\frac{1}{\Qzero}.
\end{equation}
Here, $\omegar = 2\pi \fr$ with $\fr$ the resonance frequency, $\kB$ is the Boltzmann constant, $T$ is the sample temperature, $\nc$ is the characteristic saturation photon number, $\beta$ is the saturation exponent, and $\Qzero$ represents other power-independent loss. Fits were performed independently for each resonator.

\subsection{Temperature dependent loss modelling}
\label{method:tdep}
Temperature-dependent $\Qi$ data were modelled using different descriptions depending on the observed response.
For the optimized films, the $\Qi$ is well described using a standard sum of TLS, quasiparticle (QP), and other contributions,
\begin{equation}\label{eq:QiTemperature}
\frac{1}{\Qi(T)}=\frac{1}{Q_{\mathrm{TLS}}(\navg,T)}+\frac{1}{Q_{\mathrm{QP}}(T)}+\frac{1}{\Qother}.
\end{equation}
\subsubsection*{TLS contribution}
The TLS contribution is given by~\cite{Crowley23}
\begin{multline}\label{eq:QTLS}
\frac{1}{Q_{\mathrm{TLS}}(\navg, T)} = \\
\frac{1}{\Qtlsz}\,
\frac{\tanh\!\left(\dfrac{\hbar\omegar}{2\kB T}\right)}{\sqrt{1+\dfrac{\navg^{\beta_2}}{D\,T^{\beta_1}}
\tanh\!\left(\dfrac{\hbar\omegar}{2\kB T}\right)}}
,
\end{multline}
where $D$ is a scaling parameter, and $\beta_1$ and $\beta_2$ are empirical exponents.

\subsubsection*{Quasiparticle contribution and frequency shift}

Thermally excited quasiparticles (QPs) contribute to both the loss and the reactance of superconducting resonators.
We describe both contributions using Mattis--Bardeen theory~\cite{Mattis58,Gao08,Zmuidzinas12}
in the limit $\kB T \ll \Delta_0$ and $\hbar\omegar \ll \Delta_0$, assuming a thermal-equilibrium QP density.
Here $\Delta_0 = 1.764\,\kB\Tc$, with $\Tc$ the superconducting critical temperature.

In this limit, the QP-limited quality factor is~\cite{McCarrick14}
\begin{equation}
\frac{1}{Q_{\mathrm{QP}}(T)} =
\frac{4\alpha_\mathrm{k}^{(Q)}}{\pi}\,
e^{-\Delta_0/\kB T}
\sinh(\xi)\,K_0(\xi),
\label{eq:Q_qp}
\end{equation}
where $\xi = \hbar\omegar/2\kB T$ and $K_0$ is the modified Bessel function of the second kind.
The corresponding fractional frequency shift is~\cite{McCarrick14}
\begin{equation}
\frac{\delta f(T)}{\fzero} =
-\frac{\alpha_\mathrm{k}^{(f)}}{2}\,
S_2\,
\sqrt{\frac{2\pi\kB T}{\Delta_0}}\,
e^{-\Delta_0/\kB T},
\label{eq:df_qp}
\end{equation}
with
\begin{equation}
S_2 = 1 + \sqrt{\frac{2\Delta_0}{\pi\kB T}}\,e^{-\xi} I_0(\xi),
\label{eq:S2}
\end{equation}
where $I_0$ is the modified Bessel function of the first kind,
$\delta f(T) = f(T) - \fzero$, and $\fzero$ is the resonance frequency at the lowest measured temperature.
In Mattis--Bardeen theory, both expressions contain the same kinetic-inductance fraction $\alpha_\mathrm{k}$,
which we extract independently from the loss ($\alpha_\mathrm{k}^{(Q)}$) and from the frequency shift ($\alpha_\mathrm{k}^{(f)}$).

\subsubsection*{Thermally activated dissipation model}

For devices exhibiting a degradation of $\Qi$ with increasing temperature, the excess dissipation was quantified by subtracting the lowest-temperature loss and fitting the resulting increase in inverse quality factor to a standard Arrhenius form~\cite{Bahrami26,Coffey91,Vinokur90,Blatter94,Song09,Silva17},
\begin{equation}
\delta\!\left(\frac{1}{Q}\right)
=
\frac{1}{\Qi(T)}-\left(\frac{1}{\Qi}\right)_0
=
A_v \exp\!\left(-\frac{\Tact}{T}\right),
\end{equation}
with prefactor $A_v$ and activation temperature $\Tact$.
The corresponding frequency shift was analyzed using the same $\Tact$ and
\begin{equation}
\frac{\delta f(T)}{\fzero}
=
A_f \exp\!\left(-\frac{\Tact}{T}\right),
\end{equation}
with prefactor $A_f$ as only free parameter. Fitting details and results are provided in the Supplementary Information.

\clearpage

\section*{Data and code availability}

All raw and processed data and figure materials supporting this study, together with the code used for data analysis, modeling, and figure generation, are available at \url{https://doi.org/10.4121/ab78c925-d3c2-4bc5-929c-39a85db7786e}.

\section*{Acknowledgements}
We thank S.~Vallés-Sanclemente for assistance with measurements; M.~Fischer and L.~Roure for support with sputter systems;
F.~Stallone for support with annealing of sapphire wafers; G.~Steele for sharing PCBs and sample-holder design; Kavli Nanolab Delft staff for cleanroom support; and G.~Scappucci and M.~Fischer for helpful discussions. We acknowledge funding from the European Union Flagship on Quantum Technology (OpenSuperQplus100, No. 101113946), the Dutch National Growth Fund (KAT-1), Intel Corporation, and the Allowance for Top Consortia for Knowledge and Innovation (TKIs) of the Dutch Ministry of Economic Affairs.

\section*{Author contributions}
A.K. conceptualized the project. M.B., M.F., and A.K. designed the resonator devices and K.K.T. calculated participation ratios. A.K. fabricated the resonator devices and optimized fabrication recipes. H.M. acquired the X-ray diffraction data. A.K. acquired AFM images. K.K.T. designed and fabricated the qubit device. L.H. and A.K. performed the measurements using dilution and $\Hethree$ refrigerators, respectively. A.K. developed the data-analysis methodology. A.K. wrote the manuscript with contributions from L.H. and L.D.C., and feedback from all authors. L.D.C. supervised the project.

\section*{Competing interests}

A.K., H.M., and L.D.C. are listed as inventors on a patent application related to this work. The authors declare no other competing interests.

\clearpage


\clearpage

\onecolumn

\section*{Supplementary Information}
\vspace{1em}

\setlength{\parindent}{0pt}
\setlength{\parskip}{0.4em}

\setcounter{section}{0}
\setcounter{subsection}{0}
\setcounter{subsubsection}{0}
\setcounter{figure}{0}
\setcounter{table}{0}

\renewcommand{\thesection}{S\arabic{section}}
\renewcommand{\thesubsection}{S\arabic{section}.\arabic{subsection}}
\renewcommand{\thesubsubsection}{S\arabic{section}.\arabic{subsection}.\arabic{subsubsection}}
\renewcommand{\thefigure}{S\arabic{figure}}
\renewcommand{\thetable}{S\arabic{table}}
\renewcommand{\theHsection}{S\arabic{section}}
\renewcommand{\theHsubsection}{S\arabic{section}.\arabic{subsection}}
\renewcommand{\theHsubsubsection}{S\arabic{section}.\arabic{subsection}.\arabic{subsubsection}}

\renewcommand{\theHfigure}{S\arabic{figure}}
\renewcommand{\theHtable}{S\arabic{table}}

\section{Parameters of film growth}
\label{supsec:depo_details}
To map the growth window for $\aTa$, we varied the Ar sputter flow and Ti seed-layer thickness at room temperature and compared the resulting film phase across two sputter systems. We also include high-temperature $\aTa$ reference films grown without a seed layer and a representative Nb-seeded film grown at room temperature. The deposition conditions and dominant phase identified for all films grown are summarized in Table~\ref{tab:growth_parameters}.

\begin{sidewaystable}[htbp]

\caption{\textbf{Summary of film deposition parameters and resulting film phase and texture.}
Films are grouped according to growth series: T denotes films grown on tiles (T1--T10), while W denotes films grown on wafers (WA11--WC1). The dashed line separates the two groups. RT denotes room temperature, while RF and DC denote radio-frequency and direct-current sputtering modes, respectively. Seed-layer thicknesses are calculated from the calibrated growth rate and deposition time, corrected for shutter opening and closing. The assignment of dominant phase is based on XRD and/or AFM characterization.}

\label{tab:growth_parameters}

\centering
\setlength{\tabcolsep}{3pt}
\renewcommand{\arraystretch}{1.15}

\begin{tabular}{@{}cccccccccccc@{}}
\toprule
\shortstack{Tile/\\wafer} &
\shortstack{Sputter\\unit} &
\shortstack{Process\\temp.} &
\multicolumn{5}{c}{Seed-layer parameters} &
\multicolumn{3}{c}{Ta deposition parameters (DC)} &
\shortstack{Dominant\\phase} \\
\cmidrule(lr){4-8}
\cmidrule(lr){9-11}
& & &
\shortstack{Material} &
\shortstack{Power\\$(\W)$} &
\shortstack{Flow\\$(\sccm)$} &
\shortstack{Pressure\\$(\muBar)$} &
\shortstack{Thickness\\(\AA)} &
\shortstack{Power\\$(\W)$} &
\shortstack{Flow\\$(\sccm)$} &
\shortstack{Pressure\\$(\muBar)$} &
\\
\midrule

T1   & AC450 & RT            & -- & --       & -- & --  & --  & 300 & 9  & 7   & $\bTa(002)$ \\
T2   & AC450 & RT            & -- & --       & -- & --  & --  & 300 & 13 & 7   & $\bTa(002)$ \\
T3   & AC450 & RT            & -- & --       & -- & --  & --  & 300 & 17 & 7   & $\bTa(002)$ \\
T4   & AC450 & RT            & Ti & 100 (RF) & 20 & 40  & $1.0 \pm 0.5$ & 300 & 17 & 7   & mixed $\mTa$ \\
T5   & AC450 & RT            & Ti & 100 (RF) & 20 & 40  & $2.6 \pm 0.5$   & 300 & 17 & 7   & $\aTa(110)$ \\
T6   & AC450 & RT            & Ti & 100 (RF) & 20 & 40  & $5.6 \pm 0.5$   & 300 & 17 & 7   & $\aTa(110)$ \\
T7   & AC450 & RT            & Ti & 100 (RF) & 20 & 40  & $2.6 \pm 0.5$   & 300 & 9  & 7   & $\aTa(110)$ \\
T8   & AC450 & RT            & Ti & 100 (RF) & 20 & 40  & $2.6 \pm 0.5$   & 300 & 13 & 7   & $\aTa(110)$ \\
T9   & AC450 & RT            & Ti & 100 (RF) & 20 & 40  & $2.6 \pm 0.5$   & 300 & 17 & 7   & $\aTa(110)$ \\
T10  & AC450 & RT            & Ti & 100 (RF) & 20 & 40  & $2.6 \pm 0.5$   & 300 & 21 & 7   & mixed $\mTa$ \\
\addlinespace
\hdashline
\addlinespace
WA11 & AC450 & RT            & Ti & 100 (RF) & 20 & 40  & $2.6 \pm 0.5$   & 300 & 17 & 7   & $\aTa(110)$ \\
WA12 & AC450 & RT            & Ti & 100 (RF) & 20 & 40  & $2.6 \pm 0.5$   & 300 & 13 & 7   & $\aTa(110)$ \\
WA13 & AJA   & RT            & Ti & 100 (DC) & 30 & 1.8 & $2.5 \pm 0.3$ & 178 & 50 & 2.7 & $\aTa(110)$ \\
WA14 & AJA   & RT            & Ti & 100 (DC) & 30 & 1.8 & $2.5 \pm 0.3$ & 178 & 30 & 1.8 & $\aTa(110)$ \\
WA15 & AJA   & RT            & Ti & 100 (DC) & 30 & 1.8 & $3.3 \pm 0.3$ & 178 & 30 & 1.8 & mixed $\mTa$ \\
WB1  & AJA   & $750\degC$ & -- & --       & -- & --  & --  & 178 & 50 & 1.8 & $\aTa(111)$ \\
WB2  & AJA   & $750\degC$ & -- & --       & -- & --  & --  & 178 & 50 & 1.7 & $\aTa(111)$ \\
WB3  & AJA   & $750\degC$ & -- & --       & -- & --  & --  & 178 & 50 & 1.7 & $\aTa(111)$ \\
WC1  & AJA   & RT            & Nb & 180 (DC) & 50 & 2.7   & $50 \pm 1$  & 178 & 50 & 1.7 & $\aTa(110)$ \\
\botrule
\end{tabular}
\end{sidewaystable}

\clearpage

\section{Summary of resonator devices}

Supplementary Table~\ref{tab:device_cleaning_conditions} provides a summary of the ten devices used, including device label, provenant wafer, film growth method, film texture, and cleaning method. It also indicates the dilution refrigerator used for low-temperature measurements, and whether the device showed thermally activated loss. Refrigerator details are provided in Supplementary Section~\ref{supsec:measurement_setup}.

\begin{table}[htbp]
\caption{\textbf{Summary of resonator devices.}
Dashed  lines separate devices with Ti-seeded $\aTa(110)$, unseeded high-temperature-grown $\aTa(111)$, and Nb-seeded $\aTa(110)$ films. The cleaning procedure labelled 'optimized' consists of a 20~min piranha clean followed by a 13~min dip in 7:1 BOE.}
\label{tab:device_cleaning_conditions}
\centering
\setlength{\tabcolsep}{3pt}
\renewcommand{\arraystretch}{1.15}

\begin{tabular*}{\textwidth}{@{\extracolsep{\fill}}ccccccccc@{}}
\toprule
\shortstack{Device} &
\shortstack{Wafer} &
\shortstack{Seed\\layer} &
\shortstack{Film\\texture} &
\shortstack{Piranha\\clean} &
\shortstack{BOE\\duration\\(min)} &
\shortstack{Cleaning\\category} &
\shortstack{Dilution\\refrigerator(s)} &
\shortstack{Activated\\loss\\observed?} \\
\midrule
D1 & WA12 & Ti & $\aTa(110)$ & No  & No   & Control & A1 & No \\
D2 & WA11 & Ti & $\aTa(110)$ & No  & 1    & Intermediate BOE & A1, D & No \\
D3 & WA12 & Ti & $\aTa(110)$ & No  & 2.5  & Intermediate BOE & A1, D & No \\
D4 & WA11 & Ti & $\aTa(110)$ & No  & 30   & Prolonged BOE & A1 &  Yes \\
D5 & WA13 & Ti & $\aTa(110)$ & Yes & 13   & Optimized & A2, D & No \\
\addlinespace
\hdashline
\addlinespace
D6 & WB3  & -- & $\aTa(111)$ & No  & No   & Control & C & Yes \\
D7 & WB3  & -- & $\aTa(111)$ & Yes & 13   & Optimized & C & Yes \\
D8 & WB1  & -- & $\aTa(111)$ & No  & 30   & Prolonged BOE & B & Yes \\
D9 & WB2  & -- & $\aTa(111)$ & No  & 30   & Prolonged BOE & A1, D\textsuperscript{*} & Yes \\
\addlinespace
\hdashline
\addlinespace
D10 & WC1 & Nb & $\aTa(110)$ & Yes & 13 & Optimized & C & No \\
\botrule
\end{tabular*}

\begin{tablenotes}
\item[\textsuperscript{*}] For device D9, Fridge D was used only to determine~$\Tc$.
\end{tablenotes}
\end{table}
\clearpage

\section{Layout of resonator devices and CPW geometry sweep}
\label{supsec:resonator_layout}

\begin{figure}[htbp]
\centering
\includegraphics[width=\linewidth]{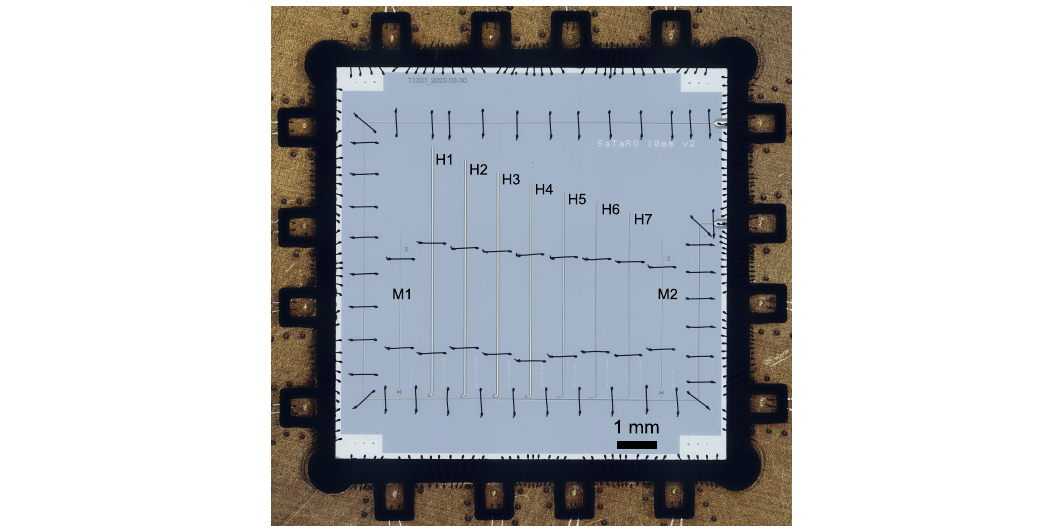}
\caption{\textbf{Optical image of a packaged resonator evice.}
Representative packaged chip showing resonators H1--H7. Two additional resonators (M1 and M2) with low design $\Qc$  serve as frequency markers and are not included in the analysis.}
\label{fig:supp_device_image}
\end{figure}

Each resonator device contains seven hanger-type CPW resonators, labelled H1 to H7, capacitively coupled to a common feedline at one end and short-circuited at the opposite end. Supplementary Fig.~\ref{fig:supp_device_image} shows an optical image of a representative packaged device after wirebonding. Crossbonds connecting ground planes are placed across both the feedline and each resonator to suppress wave propagation in slotline modes~\cite{Chen14a}.

The resonators were designed with different CPW dimensions, resonance frequencies, and coupling quality factors, as summarized in Supplementary Table~\ref{tab:resonator_geometries}. Resonators H1 to H3 have a common, largest CPW gap width ($20~\micron$) and different design $\Qc$.  Resonators H3 to H7 have progressively smaller widths, and common design $\Qc$. Devices D8 and D9 use an earlier resonator layout that differs only in design frequencies.

\begin{table}[htbp]
\caption{\textbf{Summary of resonators.}
CPW dimensions, design frequency and coupling quality factor of resonators H1 to H7 in resonator devices.
CPW dimensions are given as the center conductor width and gap width.
Design frequencies in parentheses correspond to the earlier resonator layout, used only for devices D8 and D9.}
\label{tab:resonator_geometries}
\centering
\setlength{\tabcolsep}{5pt}
\renewcommand{\arraystretch}{1.15}

\begin{tabular*}{0.9\textwidth}{@{\extracolsep{\fill}}cccc@{}}
\toprule
\shortstack{Resonator} &
\shortstack{CPW dimensions\\    centre--gap $(\micron)$} &
\shortstack{Design frequency\\  $(\GHz)$} &
\shortstack{Design $\Qc$\\     $(\times 10^6)$} \\
\midrule
H1 & 36--20   & 5.00 (5.30) & 1.8 \\
H2 & 36--20   & 5.25 (5.50) & 1.2 \\
H3 & 36--20   & 5.50 (5.90) & 0.6 \\
H4 & 28.8--16 & 5.75 (6.10) & 0.6 \\
H5 & 21.6--12 & 6.00 (6.30) & 0.6 \\
H6 & 14.4--8  & 6.25 (6.50) & 0.6 \\
H7 & 7.2--4   & 6.50 (6.70) & 0.6 \\
\botrule
\end{tabular*}
\end{table}

\clearpage
\section{Cryogenic measurement setups}
\label{supsec:measurement_setup}

In total, four setups were used for cryogenic characterization of devices. Three dilution refrigerators were used to perform measurements up to $300~\mK$: Leiden Cryogenics CF-450 (Fridge A), Leiden Cryogenics CS-81 (Fridge B), and Leiden Cryogenics MNK 650 CF (Fridge C). A $\Hethree$ dipstick refrigerator (Fridge D) was used for measurements between $250~\mK$ and $1~\Kelvin$.

\begin{figure}[htbp]
\centering
\includegraphics[width=\linewidth]{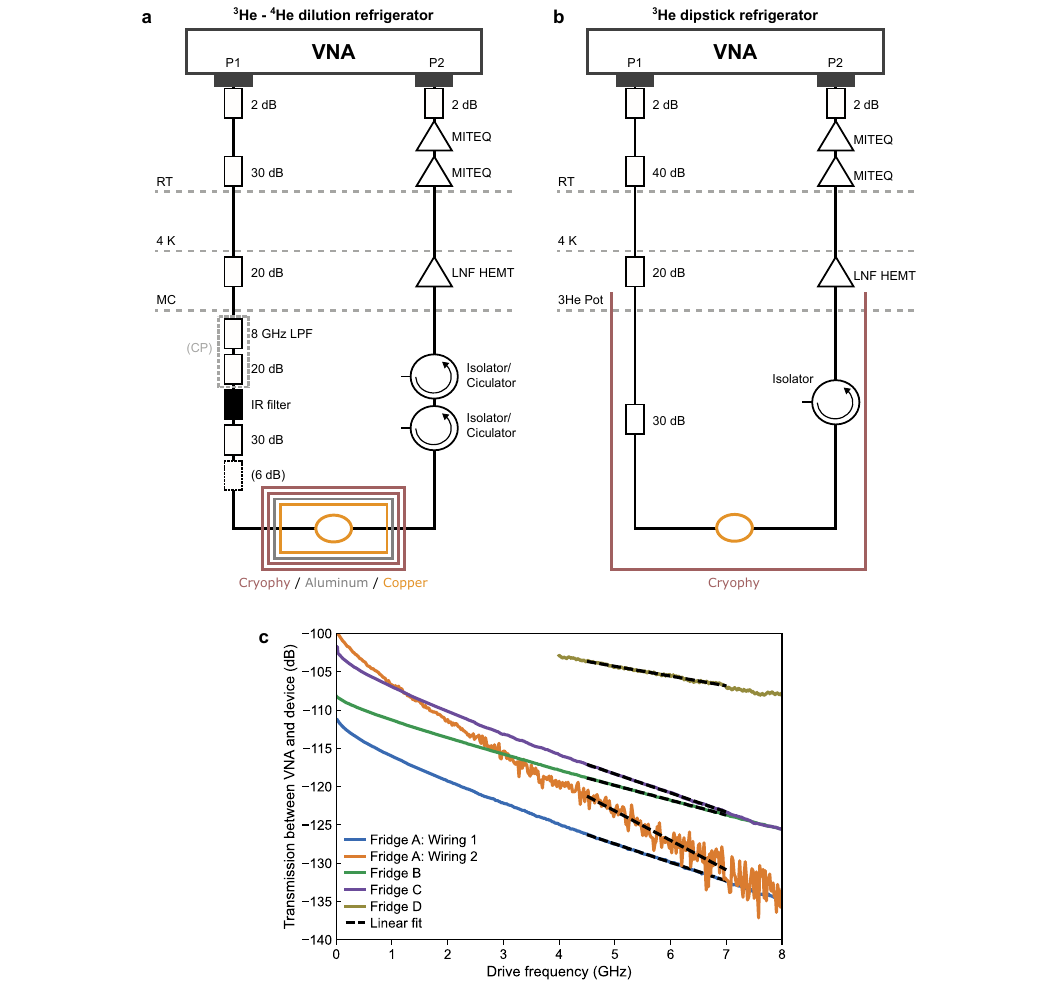}
\caption{\textbf{Measurement setups used for the cryogenic characterization of resonator devices.} (a) Wiring configuration of the three $\Hethree$--$\Hefour$ dilution refrigerators (fridges A--C) used for characterization up to $300~\mK$. The cable constructions differ in each refrigerator, but the microwave components are configured to be similar. An additional $6~\dB$ attenuator was used in fridge A wiring configuration 1. In wiring configuration 2, the $8~\GHz$ low-pass filter and $20~\dB$ attenuator are thermalized to the cold-plate stage. (b) Wiring configuration of the $\Hethree$ dipstick refrigerator used for characterization between $250~\mK$ and $1~\Kelvin$. (c) Input-line insertion loss as a function of frequency for all setups (see text for details). The dashed lines are linear fits between $4.5$ and $7~\GHz$.}
\label{fig:supp_fridge_wiring}
\end{figure}

The devices were mounted in custom copper housings attached to the cold stages of the respective refrigerators. In the dilution refrigerators, the housing was enclosed by a copper shield (internally coated with an infrared (IR) absorbing layer) followed by magnetic shields: single-layer aluminium and double-layer Cryophy~\cite{Bruno15}. In the $\Hethree$ refrigerator, the  housing was enclosed within the brass inner vacuum can (not shown in Supplementary Fig.~\ref{fig:supp_fridge_wiring}) and a single-layer Cryophy magnetic shield.

The wiring configuration includes commercial components: attenuators (XMA 2082-6243-dB-CRYO), low-pass filters (K\&L Microwave 6L250-8000/T18000-O/O), isolators and circulators (LNF: CIISC4\_12A, CIC4\_12B, Quinstar: CTH0408KCS, QCY-060400CM00, and QCI-075900XM00), cryogenic amplifiers (LNF LNC4\_8), and room-temperature amplifiers (Miteq AFS3-0400800-10-ULN and Miteq AFS3-00101200-35-ULN-R). The IR filter is homemade using eccosorb. Measurements were performed with a Rohde \& Schwarz vector network analyzer (VNA) ZNB20.

The cryogenic wiring configurations of the refrigerators are shown in Supplementary Fig.~\ref{fig:supp_fridge_wiring}a-b. Each measurement setup has variations in insertion loss (IL) due to differences in cable material and lengths, which are characterized to estimate the input power required to estimate $\navg$ using Eq.~\ref{eq:supp_photonnumber}.

The characterization of the input cables use round-trip $\Strans$ measurement, which includes all the cables from the VNA to the lowest temperature of the refrigerator. In the dilution refrigerators, these measurements were done at base temperature using two identical cable chains, and the IL taken as $|\Strans|/2$. The cable pairs used in the round-trip $\Strans$ measurement are identical to the cables used for the measurements presented in this work. The room-temperature $30~\dB$ attenuator and components at the mixing chamber were removed for these cryogenic calibrations. The IL of the attenuators is assumed to be ideal. The IL of the low-pass filter is assumed to be negligible below $8~\GHz$. The homemade eccosorb filters are assumed to have $\sim3~\dB/\GHz$ IL. The IL characterization of fridge D was performed at room temperature, which overestimates the IL of the semi-rigid coaxial cables when cold~\cite{Pobell96}. The total IL, obtained by combining the cable attenuation with the known attenuation of the microwave filters and attenuators, is shown in Supplementary Fig.~\ref{fig:supp_fridge_wiring}c. Note that two different input-line configurations were used in fridge A. While wiring 1 uses semi-rigid coaxial cables, similar to all other refrigerator wiring configurations,  wiring 2 uses flexible cables (Delft Circuits). Each transmission curve was fit linearly from $4.5$ to $7~\GHz$, and the resulting fit used to calculate the input power.

Assumptions made in the calibration of the IL contribute to systematic error in $\navg$. Variations in the IL of each microwave component are expected, especially the homemade eccosorb filters. Also, while they are nominally identical, there may be differences between cables used for calibration and measurement. These systematic offsets in $\navg$ are not expected to influence any results presented.

\clearpage

\section{Microwave transmission measurements and resonator analysis}
Microwave transmission measurements were performed using the VNA by recording the complex transmission coefficient $\Strans$ near the resonant frequency of the CPW resonators at cryogenic temperatures. A schematic of the measurement wiring and input attenuations is shown in Supplementary Fig.~\ref{fig:supp_fridge_wiring}.

The transmission response was acquired in multiple frequency segments to achieve high resolution around the resonance while maintaining a broad measurement span~\cite{Baity24}. A representative dataset is shown in Supplementary Fig.~\ref{fig:S21_complex_plot}, corresponding to the highest photon number measurement of resonator H3 in Fig.~\ref{fig:fig4}a.

\begin{figure*}[htbp]
\centering
\includegraphics[]{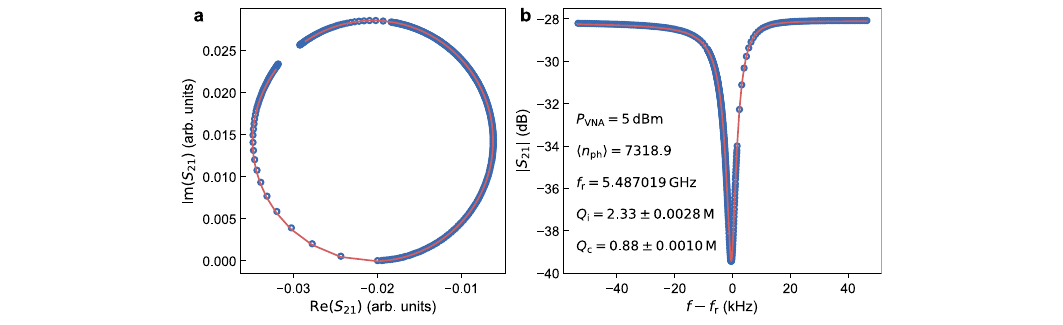}
\caption{\textbf{Representative microwave transmission measurement.}
Complex-valued transmission coefficient $\Strans$ of resonator H3 at the highest measured photon number in Fig.~\ref{fig:fig4}a. (a) Complex-plane representation of $\Strans$, showing the characteristic resonance circle. (b) Transmission magnitude $|\Strans|$ as a function of frequency. Solid curves represent fits of Eq.~\ref{eq:complex_resonator_model}.
}
\label{fig:S21_complex_plot}
\end{figure*}

The resonator response was modeled using the complex transmission expression described in Ref.~\cite{Bruno15}, which accounts for impedance mismatch, cable delay, and resonance asymmetry:
\begin{equation}\label{eq:complex_resonator_model}
\Strans(f)=A(1+B\delta \tilde{f})\left[1-\frac{\Ql}{\Qe}\frac{e^{i\theta}}{1+2i\Ql\delta \tilde{f}}\right]e^{i(\phi_{\nu}\delta \tilde{f}+\phi_0)},
\end{equation}
where $\delta \tilde{f}=(f-\fr)/\fr$ is the normalized frequency detuning from resonance, $\Ql$ is the loaded quality factor, $\Qe$ the external quality factor and $\theta$ the rotation angle describing the resonance asymmetry. Further, $A$ and $B$ account for the cable attenuation, and $\phi_{\nu}$ and $\phi_0$ account for the propagation delay of the cables.

The model was fitted simultaneously to the real and imaginary components of $\Strans$ using bounded nonlinear least-squares optimization. The coupling quality factor was obtained as $\Qc=\Qe/\cos\theta$, and the internal quality factor as $1/\Qi = 1/\Ql - 1/\Qc$.

Uncertainties in the fitted parameters were estimated from the covariance matrix obtained from the Jacobian of the fit, assuming approximately Gaussian-distributed residuals, and correspond to one standard deviation.

The average photon number in the resonator was estimated as
\begin{equation}\label{eq:supp_photonnumber}
\navg= \frac{2}{\hbar \omegar^2}\frac{\Ql^2}{\Qc}\Pin.
\end{equation}
Here, $\Pin$ is the incident power at the feedline, obtained by correcting the VNA excitation power for the total attenuation of the input line, which is specific to each refrigerator (Supplementary Fig.~\ref{fig:supp_fridge_wiring}c). Supplementary Figure.~\ref{fig:S21_complex_plot} shows an example fit and best-fit $\navg$, $\fr$, $\Qi$, and $\Qc$ values.

\section{Estimating the superconducting transition temperature}
The superconducting transition temperature $\Tc$ of the Ta films was estimated from the temperature dependence of feedline transmission. At each temperature, $|\Strans|$ was averaged between $6.8$ and $7.8~\GHz$, and subsequently normalized by the value at $3.5~\Kelvin$. For each device, we took as estimate of $\Tc$ the temperature at which $|\Strans|$ lies halfway between its value at $5$ and $3.5~\Kelvin$~\cite{Song09}.

\begin{figure}[htbp]
\centering
\includegraphics[width=\linewidth]{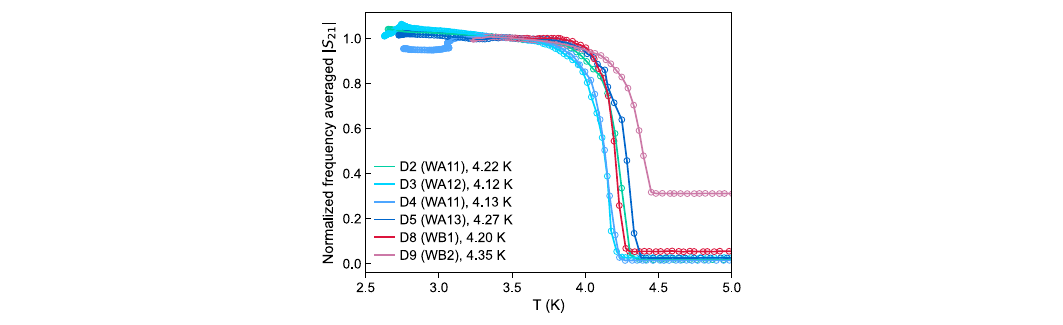}
\caption{\textbf{Temperature dependence of feedline transmission.}
Temperature dependence of the frequency-averaged, normalized feedline transmission $|\Strans|$ for various resonator devices. The temperatures reported in the legend correspond to the estimated $\Tc$.}
\label{fig:Tc}
\end{figure}
\clearpage

\section{Power dependence of $\Qi$ at $\sim20~\mK$}
\label{supsec:pow_dep_tls}

\begin{figure}[htbp]
\centering
\includegraphics[width=\linewidth]{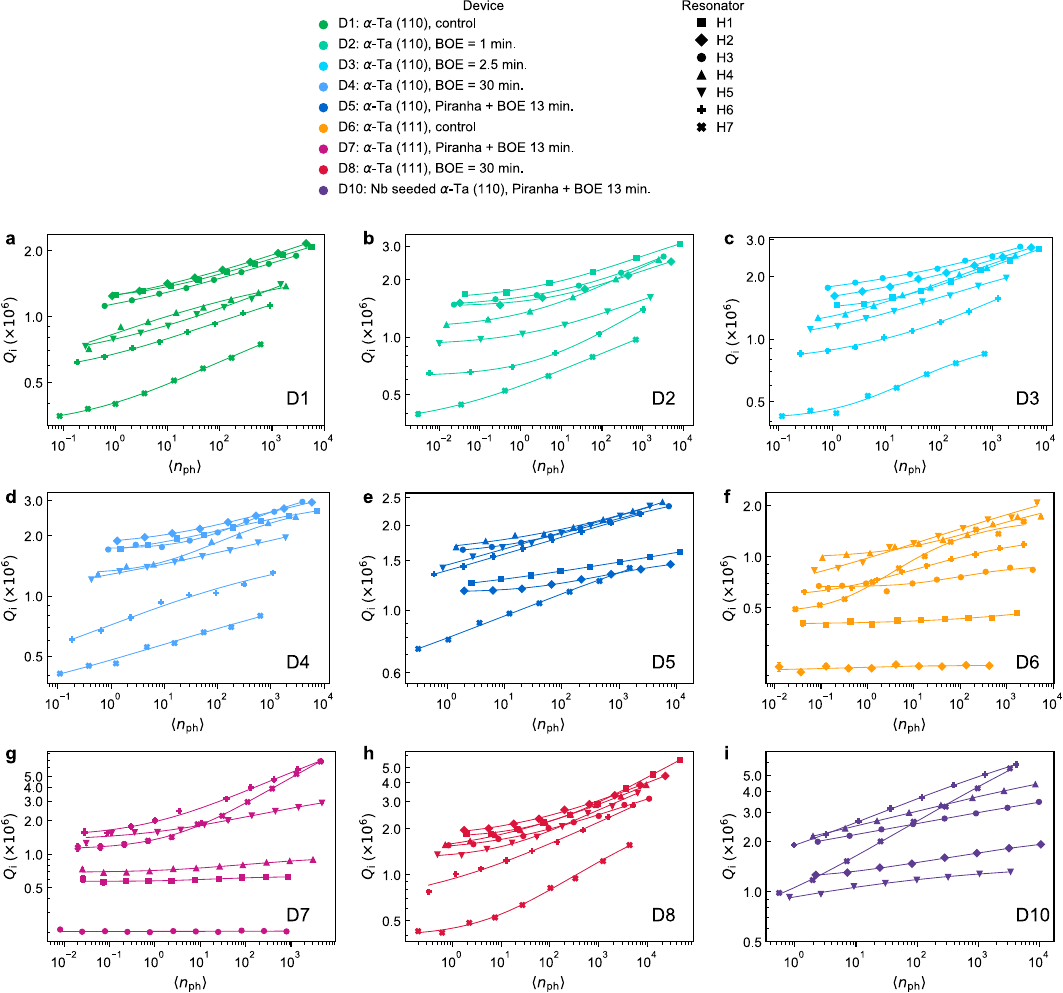}
\caption{\textbf{Power dependence of $\Qi$ at base temperature and fits of the TLS model.}
Measured internal quality factor $\Qi$ as a function of average photon number $\navg$ for all resonators across nine devices, measured at dilution refrigerator base temperature ($\sim20~\mK$). Marker shapes denote resonators H1--H7, while colors denote device and processing condition, as identified by the legend. Unless labeled Nb-seeded, $\aTa(110)$ films are Ti-seeded, and $\aTa(111)$ films are grown at high temperature without a seed layer. Solid curves are best fits of the TLS saturation model, Eq.~\ref{eq:TLSSaturation}. Panel (e) shows the same data as Fig.~\ref{fig:fig3}b.}
\label{fig:LowTPowDep}
\end{figure}

This section complements Fig.~\ref{fig:fig3}, showing (Supplementary Fig.~\ref{fig:LowTPowDep}) the measured power dependence of $\Qi$ at dilution refrigerator base temperature for all resonators across all devices, together with fits of Eq.~\ref{eq:TLSSaturation}. Best-fit parameter values are summarized in Supplementary Fig.~\ref{fig:beta_nc_model}. No base-temperature, power-dependent data were measured for device D9. Some wide-gap resonators (H1--H3) on devices D6, D7 and D10  [most notably H2 on D7 and H1 on D10 (not shown)], have $\Qi$ values that are nearly independent of power over the measured range and lower than those of comparable resonators on other devices. For these resonators, power-independent loss dominates, so $\Qtlsz$, $\beta$ and $\nc$ are not constrained by the fit.

\begin{figure}[htbp]
\centering
\includegraphics[width=\linewidth]{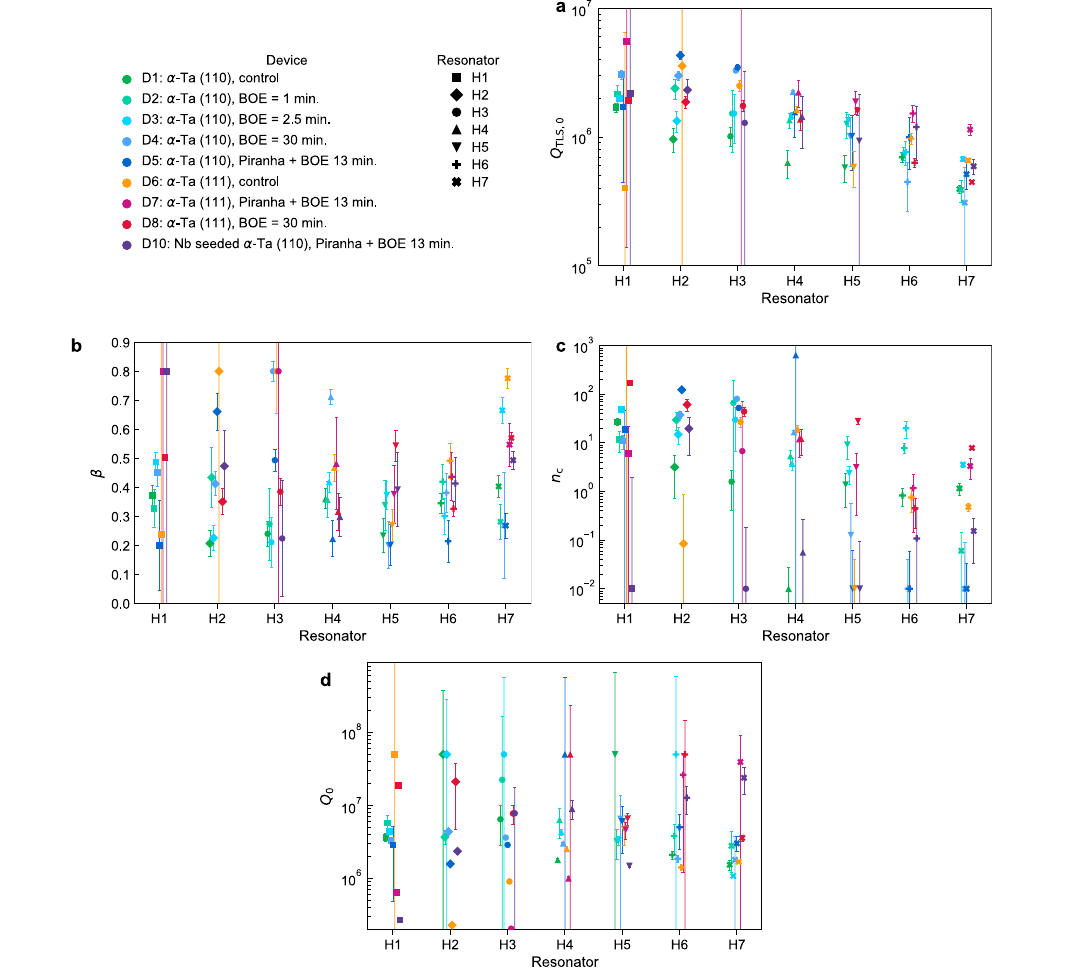}
\caption{\textbf{Summary of best-fit parameter values.}
Parameter values obtained by fitting Eq.~\ref{eq:TLSSaturation} to all data in Supplementary Fig.~\ref{fig:LowTPowDep}: (a) low-power TLS-limited quality factor $\Qtlsz$; (b) saturation exponent $\beta$; (c) characteristic photon number $\nc$; and (d) power-independent quality factor $\Qzero$. The four parameters are correlated in Eq.~\ref{eq:TLSSaturation}, and for many resonators the onset of TLS saturation $\nc$ is not well resolved within the measured power range. This results in large uncertainties of the extracted best-fit parameters.}
\label{fig:beta_nc_model}
\end{figure}
\clearpage

\section{Estimation of participation ratio}
\label{supsec:PR}
Resonators H1--H7 (Supplementary Table~\ref{tab:resonator_geometries}) were modeled in CST Studio Suite. The model included 200 $\nm$ Ta film on a $0.5~\mm$ sapphire substrate, with $2~\mm$ of vacuum above the metal, and $2~\mm$ below the substrate. For simplicity, the seed layer was neglected in the simulation.

Eigenmode simulations were performed to obtain the fundamental mode and the corresponding electric-field distribution for each resonator geometry. Following Ref.~\cite{Wenner11}, we estimated the metal-substrate (MS) participation ratio $\pMS$ from the electric field sampled at the MS interface. For the interfacial dielectric layer, we assume thickness $t_{\mathrm{MS}} = 3~\nm$ and relative dielectric constant $\epsilon_{r,\mathrm{MS}} = 10$~\cite{Wenner11,Wang15,Calusine18}.

The TLS loss is given by the participation-weighted sum over the metal--substrate (MS), substrate--air (SA) and metal--air (MA) interfaces, and the bulk substrate (sub)~\cite{Calusine18,Woods19}:
\begin{equation}
  \frac{1}{\Qtlsz} = \sum_k p_k \tan\delta_k
  = \pMS\tan\delta_\mathrm{MS} + p_\mathrm{SA}\tan\delta_\mathrm{SA}
  + p_\mathrm{MA}\tan\delta_\mathrm{MA} + p_\mathrm{sub}\tan\delta_\mathrm{sub}.
  \label{eq:PR_sum}
\end{equation}
In CPW geometries, the interface participation ratios scale proportionally with geometry, while $p_\mathrm{sub}$ remains nearly constant~\cite{Wenner11,Wang15,Murray18,Martinis22,Calusine18}:
\begin{equation}
  c_\mathrm{SA} = \frac{p_\mathrm{SA}}{\pMS} \approx \text{const.}, \quad
  c_\mathrm{MA} = \frac{p_\mathrm{MA}}{\pMS} \approx \text{const.}, \qquad
  p_\mathrm{sub} \approx \text{const.}
  \label{eq:PR_scaling}
\end{equation}
Neglecting the substrate contribution for now, Eq.~\ref{eq:PR_sum} reduces to,
\begin{equation}
  \frac{1}{\Qtlsz} = \pMS\tandeff, \qquad
  \tandeff = \tan\delta_\mathrm{MS} + c_\mathrm{SA}\tan\delta_\mathrm{SA} + c_\mathrm{MA}\tan\delta_\mathrm{MA}.
  \label{eq:tand_eff}
\end{equation}

The extracted $\pMS$ for each resonator geometry are shown in Supplementary Fig.~\ref{fig:PR}a. These values were then used to compare $\Qtlsz$ across resonator geometries and cleaning methods, as shown in Supplementary Fig.~\ref{fig:PR}b--j. The effective loss tangent $\tandeff$ (Supplementary Fig.~\ref{fig:PR}k) was obtained from a weighted least-squares fit (minimizing $\chi^2=\sum_i (r_i/\sigma_i)^2$, with residuals $r_i$ and errors $\sigma_i$ of $Q_{\mathrm{TLS},0,i}$) of Eq.~\ref{eq:tand_eff}, with the uncertainty of $\tandeff$ scaled by $\sqrt{\chi^2_\mathrm{red}}$ when $\chi^2_\mathrm{red}>1$~\cite{Birge32}, where $\chi^2_\mathrm{red}=\chi^2/(N-1)$ for $N$ resonators.

To account for the substrate contribution, we performed a separate fit (not shown) including a geometry-independent offset, $1/\Qtlsz = \pMS\tandeff + \delta_0$, with $\delta_0 = p_\mathrm{sub}\tan\delta_\mathrm{sub}$~\cite{Calusine18,Woods19}. The resulting $\delta_0$ differ between devices beyond their uncertainties and take both signs, despite the common substrate. They are therefore inconsistent with a geometry-independent bulk loss and instead reflect device-specific deviations amplified by the extrapolation to $\pMS = 0$. Moreover, combined over all devices, $\delta_0$ is consistent with zero and is bounded by $\delta_0 < 0.8\times10^{-7}$; however, its precise determination requires further study. Therefore, in Supplementary Sec.~\ref{supsec:century}, we use $\tandeff\approx 1.64\times10^{-3}$, obtained from Eq.~\ref{eq:tand_eff} for the optimally cleaned surface of device D5 (Piranha + BOE 13~min), to estimate the TLS-limited qubit $\Tone$.

\begin{figure}[htbp]
\centering
\includegraphics[width=\linewidth]{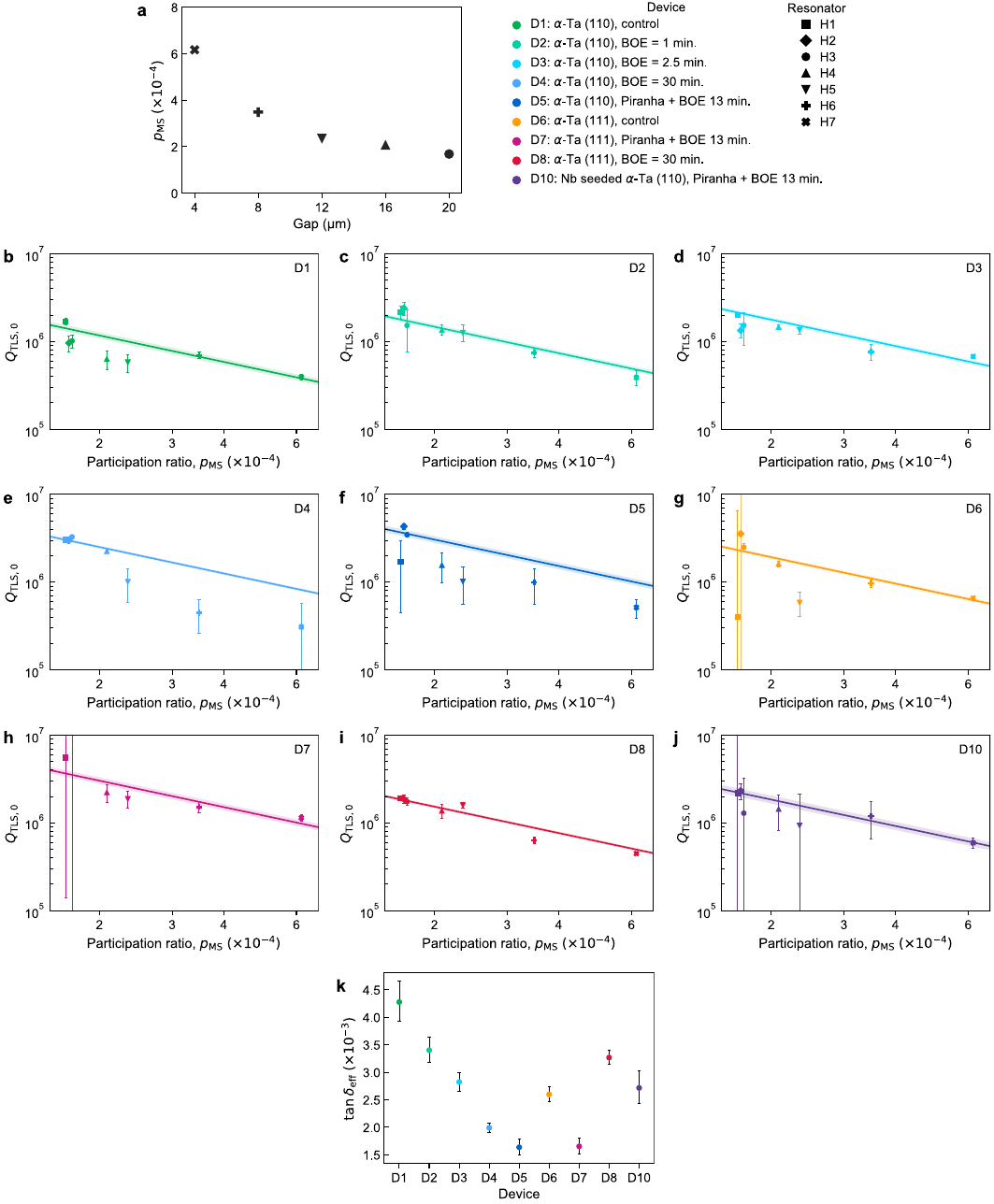}
\caption{%
(a) Simulated metal--substrate participation ratio, $\pMS$, as a function of resonator gap width for the geometry series H3--H7. The legend on the right identifies devices (colors) and resonators (markers).
(b--j) Extracted $\Qtlsz$ as a function of $\pMS$. Points with the same nominal $\pMS$ are slightly offset horizontally for visual clarity. Error bars are the uncertainties of $\Qtlsz$ from the power-dependence fits. $\Qtlsz$ increases with decreasing participation ratio, consistent with surface-dominated TLS loss. Lines are weighted fits of Eq.~\ref{eq:tand_eff} and shaded bands show the 68\% confidence interval.
(k) $\tandeff$ for each device, with 68\% confidence intervals.
}
\label{fig:PR}
\end{figure}
\clearpage

\section{Temperature and power dependence of $\aTa(110)$ resonators at moderate temperature}
\label{supsec:temp_pow_fit}
Complementing Fig.~\ref{fig:fig4}a, this section presents temperature and power dependence of extracted $\Qi$ for $\aTa(110)$ resonators across multiple devices, measured in a dilution refrigerator. Data and fits of Eq.~\ref{eq:QiTemperature} with only TLS and other contributions (i.e., ignoring QP contribution) are shown in Supplementary Fig.~\ref{fig:Qi_pow_temp}, Best-fit parameter values are shown in Supplementary Fig.~\ref{fig:beta1_beta2_model}.

\begin{figure}[htbp]
\centering
\includegraphics[width=\linewidth]{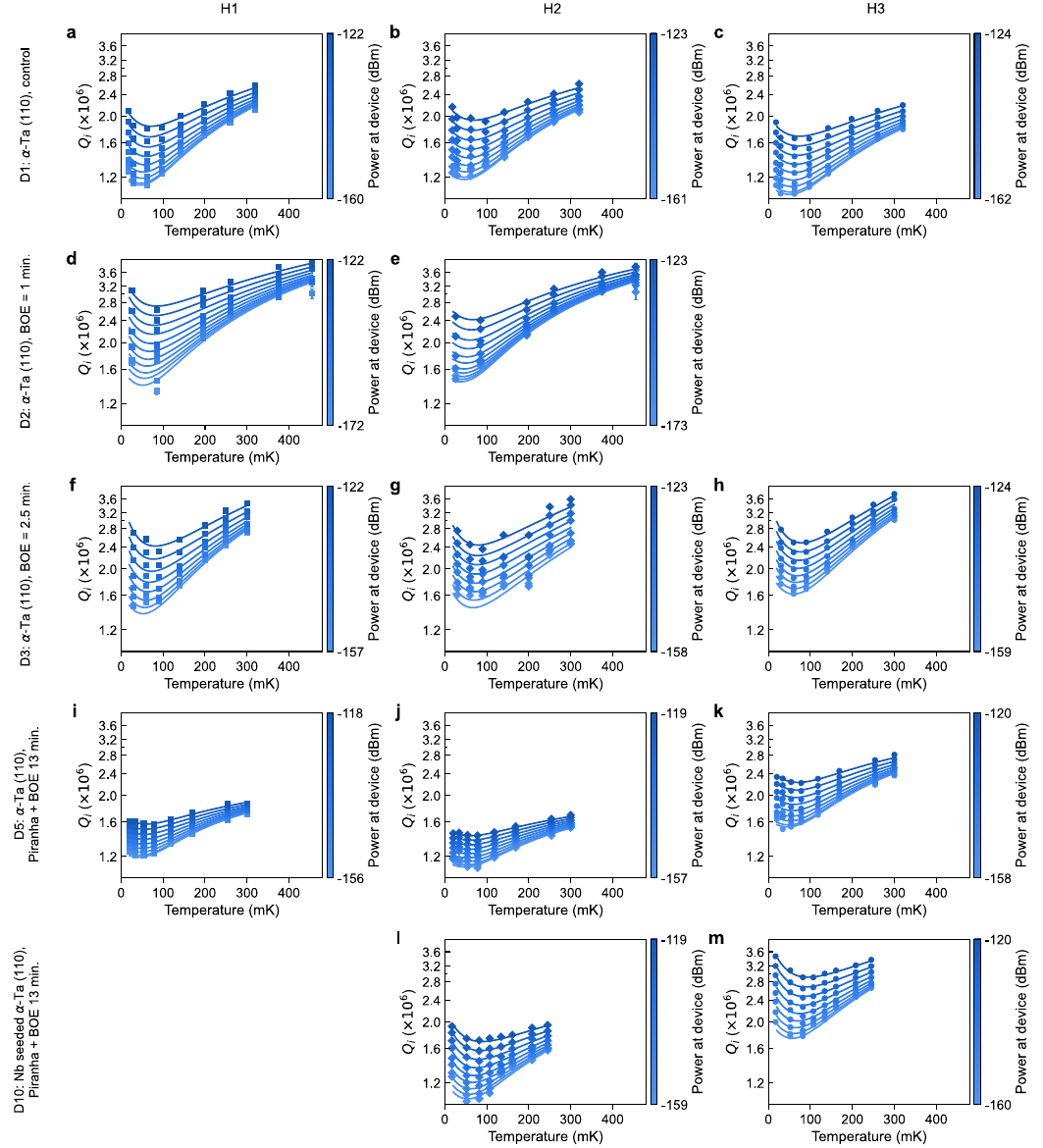}
\caption{\textbf{Temperature- and power-dependent $\Qi$ of $\aTa(110)$ resonators at moderate temperature.}
Extracted internal quality factor $\Qi$ of $\aTa(110)$ resonators as a function of temperature at different drive powers, measured in a dilution refrigerator. Each row corresponds to a specific device (indicated by text on the left side) and each column coressponds to a specific resonator (H1-H3). The device set includes Ti-seeded films subjected to different post-processing conditions and a representative Nb-seeded comparison film. Solid curves show best fits of Eq.~\ref{eq:QiTemperature} without QP contribution. Panel (h) shows the same data and fits as Fig.~\ref{fig:fig4}a. Missing panels correspond to resonator H3 of device D2, which was not measured, and resonator H1 of device D10, which is omitted because $\Qi<3\times10^{5}$ showed no systematic dependence on drive power or temperature, indicating that loss was dominated by $\Qother$.}
\label{fig:Qi_pow_temp}
\end{figure}

\begin{figure}[htbp]
\centering
\includegraphics[width=\linewidth]{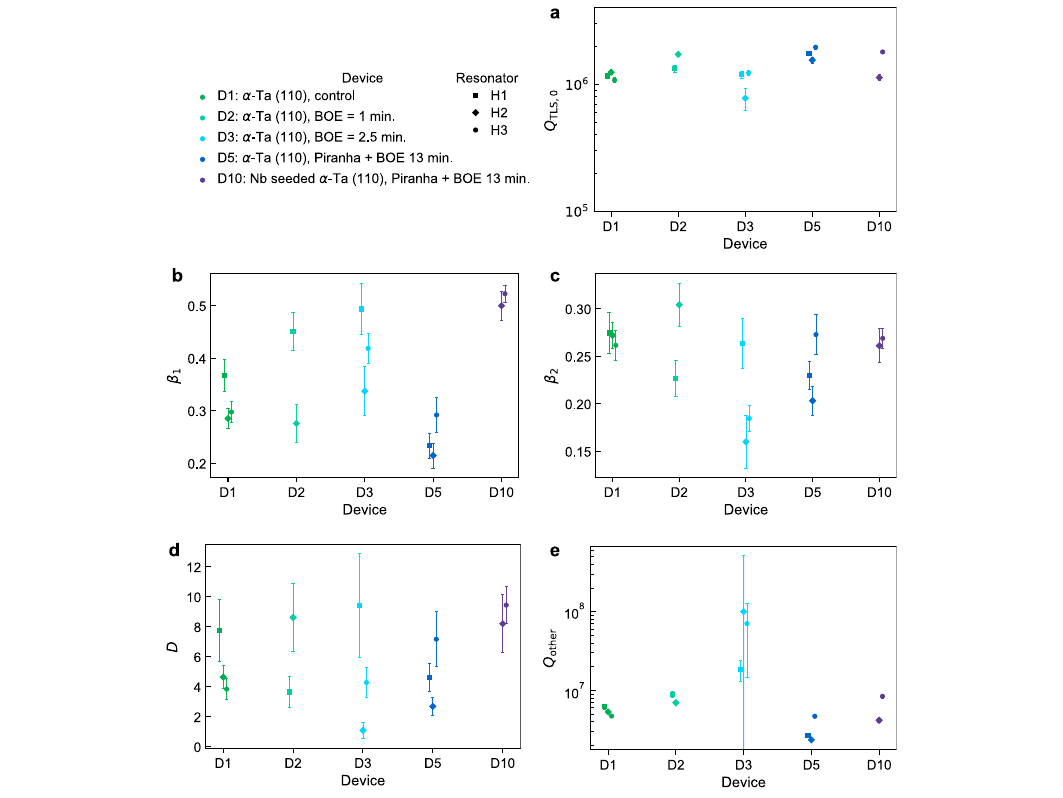}
\caption{\textbf{Summary of best-fit parameter values.}
Parameter values obtained by fitting Eq.~\ref{eq:QiTemperature} without QP contribution to data in
Supplementary Fig.~\ref{fig:Qi_pow_temp}:
(a) TLS-limited quality factor $\Qtlsz$;
(b) Exponent $\beta_1$;
(c) Exponent $\beta_2$;
(d) Saturation parameter $D$;
(e) $\Qother$.
}\label{fig:beta1_beta2_model}
\end{figure}
\clearpage

\section{Temperature and power dependence of $\aTa(110)$ resonators up to $1~\Kelvin$}
\label{supsec:110_vespa}
Complementing Fig.~\ref{fig:fig4}b-c, this section presents temperature and power dependence of extracted $\Qi$ and fractional frequency shift $\delta f/\fzero$ for multiple $\aTa(110)$ resonators across multiple devices, measured in the $\Hethree$ refrigerator up to $1~\Kelvin$.
Importantly, we include the QP contribution [Eq.~\ref{eq:Q_qp}] when fitting Eq.~\ref{eq:QiTemperature} to $\Qi$ data in this temperature range, fixing $\Tc$ to the estimated value (Supplementary Fig.~\ref{fig:Tc}).
Furthermore, since loss becomes QP dominated and also the TLS saturation term is strongly suppressed by $\tanh(\hbar\omegar/2 \kB T)$, $\beta_1$ is not constrained by these data. Therefore, we fix it for each resonator to the value extracted at lower temperatures (Supplementary Fig.~\ref{fig:beta1_beta2_model}).
We also fix $\Tc$ when fitting Eqs.~\ref{eq:df_qp} and~\ref{eq:S2} to the corresponding $\delta f/\fzero$ data at the lowest drive power. The TLS contribution to $\delta f$ is neglected; because the QP shift decreases exponentially with decreasing temperature, $\alpha_\mathrm{k}^{(f)}$ is determined by the highest-temperature data.

The kinetic-inductance fractions extracted independently from the loss and from the frequency shift,
$\alpha_\mathrm{k}^{(Q)}$ and $\alpha_\mathrm{k}^{(f)}$, lie between 2.4\% and 3.0\% and agree within 10\% for measured $\aTa(110)$ resonators (Supplementary Fig.~\ref{fig:tdep_vespa_params}e). This indicates that both the excess loss and the frequency shift at the highest temperatures are described by thermal-equilibrium QPs with the measured $\Tc$, without additional thermally activated loss mechanisms such as those reported for some $\alpha$-Ta resonators~\cite{Crowley23, Bahrami26}.

\begin{figure}[htbp]
\centering
\includegraphics[width=\linewidth]{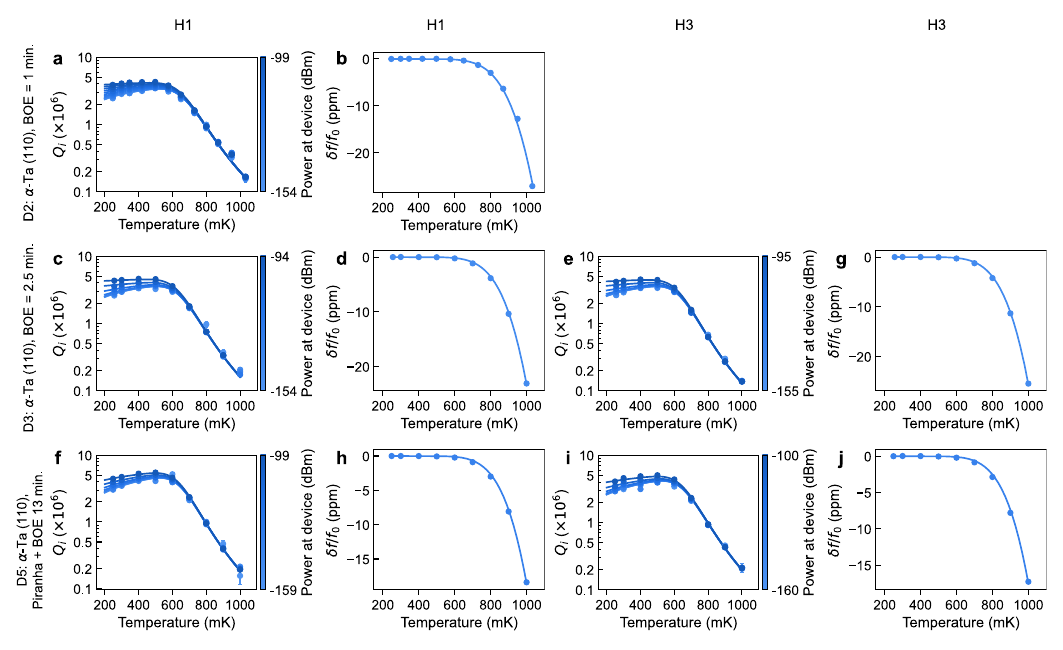}
\caption{\textbf{Temperature- and power-dependent $\Qi$ and $\delta f/\fzero$ of $\aTa(110)$ resonators up to $1~\Kelvin$.}
Extracted internal quality factor $\Qi$ and fractional frequency shift $\delta f/\fzero$ of $\aTa(110)$ resonators as a function of temperature at different drive powers, measured in the $\Hethree$ refrigerator. Each row corresponds to a specific device (indicated by text on the left side) and each column corresponds to a specific resonator (H1 or H3). The device set includes Ti-seeded films subjected to different cleaning conditions. Solid curves are best fits of Eq.~\ref{eq:QiTemperature}, now including the QP contribution [Eq.~\ref{eq:Q_qp}], and of Eqs.~\ref{eq:df_qp} and~\ref{eq:S2}. Panels (i,j) show the same data and fits as Fig.~\ref{fig:fig4}b,c. No measurements were performed on resonator H3 of device D2.
}
\label{fig:tdep_vespa}
\end{figure}

\begin{figure}[htbp]
\centering
\includegraphics[width=\linewidth]{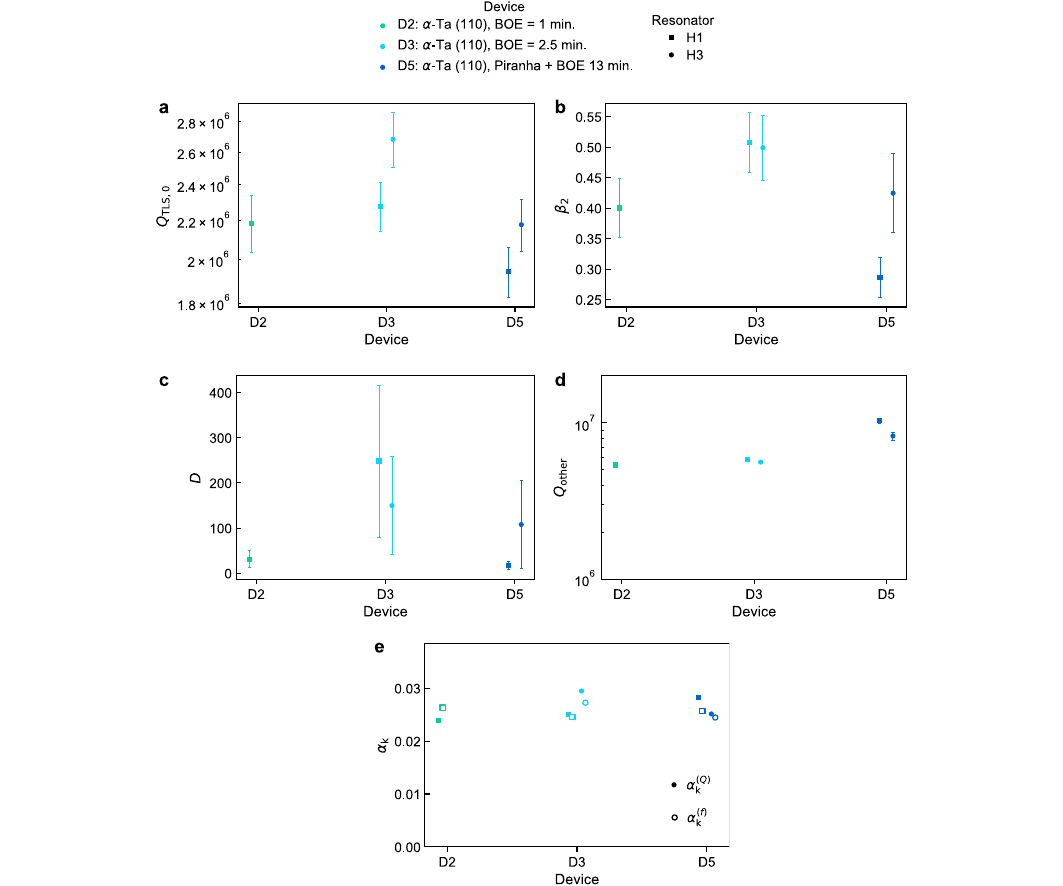}
\caption{\textbf{Summary of best-fit parameter values.}
Parameter values obtained by fitting Eq.~\ref{eq:QiTemperature} with QP contribution, and Eqs.~\ref{eq:df_qp} and~\ref{eq:S2}, to data in Supplementary Fig.~\ref{fig:tdep_vespa}:
(a) TLS-limited quality factor $\Qtlsz$;
(b) Exponent $\beta_2$;
(c) Saturation parameter $D$;
(d) $\Qother$;
(e) Kinetic-inductance fraction extracted from $\Qi$, $\alpha_\mathrm{k}^{(Q)}$ (filled symbols), and from $\delta f/\fzero$, $\alpha_\mathrm{k}^{(f)}$ (open symbols).
}
\label{fig:tdep_vespa_params}
\end{figure}

\clearpage

\section{Thermally activated excess loss and frequency shift}
\begin{figure}[htbp]
\centering
\includegraphics[width=\linewidth]{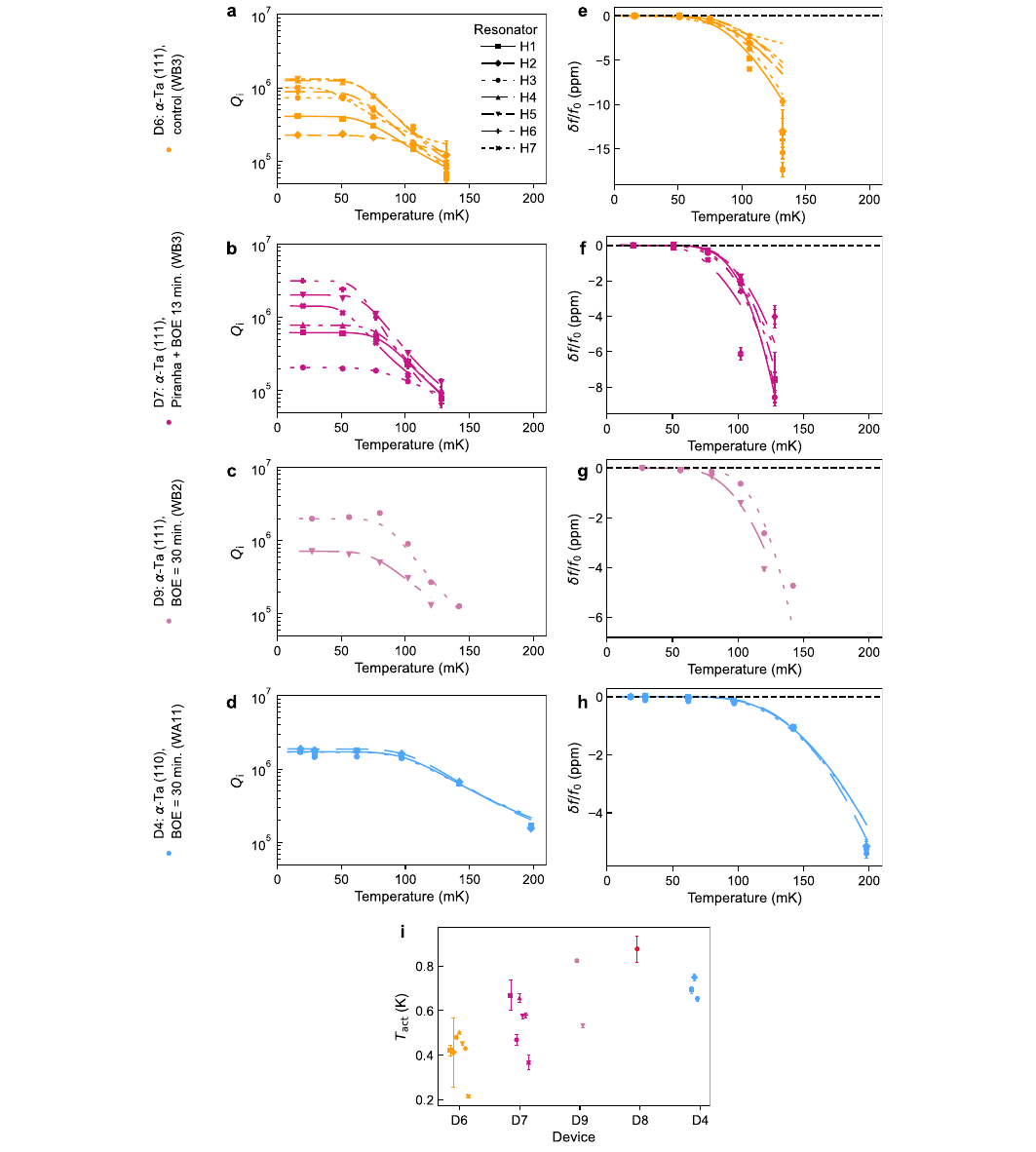}
\caption{\textbf{Thermally activated excess loss and frequency shift.}
(a--d) Temperature dependence of the internal quality factor $\Qi$ for resonators exhibiting temperature activated loss. Solid curves show fits of the thermally activated dissipation model described in Methods. The text to the left of each row identifies the device and its processing condition, while the legend in (a) defines the marker and line styles used to distinguish the individual resonators. (e--h) Corresponding fractional frequency shift $\delta f/\fzero$ and fits. Panels (a--c) and (e--g) correspond to devices with high-temperature-grown $\aTa(111)$ films and different cleaning conditions. Panels (d) and (h) correspond to a device with Ti-seeded $\aTa(110)$ film and subjected to extended buffered-oxide treatment. (i) Summary of the extracted activation temperature $\Tact$ for all datasets.}

\label{fig:tdep_dissipation}
\end{figure}

This section complements Fig.~\ref{fig:fig4}d,e, showing more examples (Supplementary Fig.~\ref{fig:tdep_dissipation}) of resonators exhibiting anomalous temperature-dependent behavior not captured by the TLS and QP models. Thermally activated loss and frequency shifts are observed in five devices: the four devices with high-temperature-grown $\aTa(111)$ and one device with Ti-seeded $\aTa(110)$ subjected to extended buffered-oxide treatment. Temperature-dependent low-power $\Qi$ and $\delta f/\fzero$ are analyzed for each resonator using the thermally activated loss model described in Methods.

\clearpage

\section{Comparison of $\Qtlsz$ across extraction methods}
\label{supsec:Qtls0_comparison}

\begin{figure}[htbp]
\centering
\includegraphics[width=\linewidth]{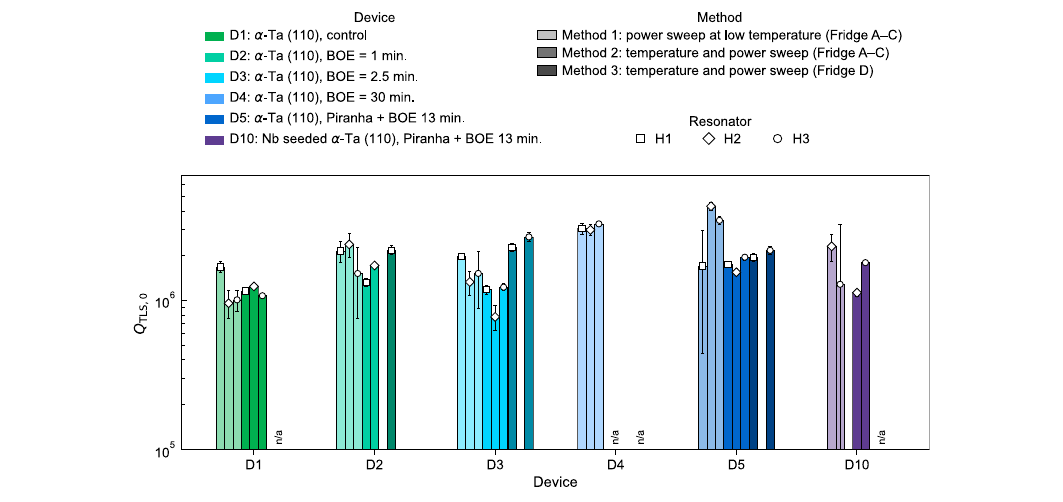}
\caption{\textbf{Comparison of $\Qtlsz$ across extraction methods.}
The legend identifies the film type and post-processing condition associated with each device. Unless labeled Nb-seeded, $\aTa(110)$ films are Ti-seeded. For each device, Methods~1--3 are shown as adjacent groups in light, base and dark shades of the device colour. Within each group, resonators H1, H2 and H3 occupy fixed positions from left to right. Empty positions indicate no measurement or excluded resonators, and ``n/a'' indicates no data for that method. Error bars show the fit uncertainties.}
\label{fig:Qtls0_comparison}
\end{figure}

We extracted $\Qtlsz$ with three methods. Method~1 uses a power sweep at low temperature in Fridges A--C, fitted with Eq.~\ref{eq:TLSSaturation}. Method~2 uses a temperature and power sweep in the same fridges, fitted with Eq.~\ref{eq:QiTemperature} using only the TLS term, Eq.~\ref{eq:QTLS}. Method~3 uses a temperature and power sweep in Fridge D, fitted with Eq.~\ref{eq:QiTemperature} including both TLS and quasiparticle terms. Only $\aTa$ (110) films are compared, since the activated behaviour of the (111) devices prevents a reliable separation of the TLS contribution. Resonator H1 of D10 is excluded, as its $\Qi$ is limited by $\Qother$. The extracted $\Qtlsz$ values are comparable across all methods (Fig.~\ref{fig:Qtls0_comparison}), with Method~1 yielding slightly higher values than Method~2.

\clearpage

\section{Measurement of transmons}
\label{supsec:century}
\begin{figure}[htbp]
\centering
\includegraphics[width=\linewidth]{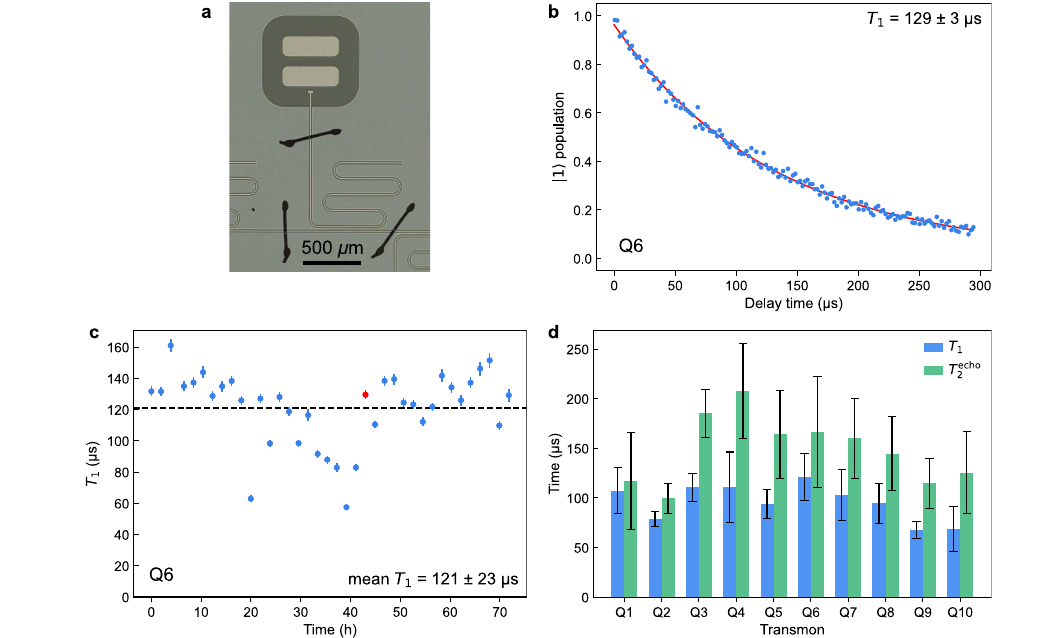}
\caption{
\textbf{Qubit device and coherence characterization.}
\textbf{a,} Optical micrograph of a representative device region, showing a single transmon capacitively coupled to a dedicated readout resonator, which itself couples to a common feedline.
\textbf{b,} Individual $\Tone$ measurement for transmon Q6, giving best-fit $\Tone=129\pm3~\us$.
\textbf{c,} Time evolution of repeated $\Tone$ measurements of Q6 over $\sim74$~h. Error bars show fit uncertainty. The red data point corresponds to the measurement in b. The dashed horizontal line indicates the time average.
\textbf{d,} Average and standard deviation of $\Tone$ and $\Techo$ for the ten transmons, measured over $\sim74$~h.
}
\label{fig:century}
\end{figure}

To assess the compatibility of Ti-seeded $\aTa(110)$ with qubits, we have designed and fabricated a test device with ten transmons with $\aTa$ capacitor pads. The transmons are designed to be minimally connected in an effort to reduce additional loss channels (Supplementary Fig.~\ref{fig:century}a). Qubit control and readout are performed through a common feedline connecting to dedicated $\lambda/4$ readout resonators. Supplementary Table~\ref{tab:century} presents design and measured values of qubit quantities of interest.

\begin{table}[htbp]
\caption{\textbf{Transmon device parameters.} Summary of designed and measured parameters of the ten-transmon device.}
\label{tab:century}
\centering
\setlength{\tabcolsep}{5pt}
\renewcommand{\arraystretch}{1.15}

\begin{tabular*}{0.9\textwidth}{@{\extracolsep{\fill}}cccc@{}}
\toprule
\shortstack{Parameter name} &
\shortstack{Parameter symbol} &
\shortstack{Designed value} &
\shortstack{Measured value} \\
\midrule
Qubit frequency $(\GHz)$ & $\omegaq/2\pi$ & 4--5.5 & 3.7--5.3 \\
Qubit-resonator coupling rate $(\MHz)$ & $g/2\pi$ & 62.4 & 47--69 \\
Resonator linewidth (kHz) & $\kappa/2\pi$  & 600 & 310--830 \\
Qubit-resonator detuning $(\GHz)$ & $\Delta/2\pi$ & 2.0--2.2 & 2.2--2.5 \\
Purcell-limited relaxation time $(\us)$ & $\TPurcell$  & 175--211 & 240--1200 \\
Relaxation time $(\us)$ & $\Tone$  &  & 68--121 \\
Hahn echo time $(\us)$ & $\Techo$  &  & 99--208 \\
\botrule
\end{tabular*}
\end{table}

The base layer, including the transmon capacitor pads, readout resonators and feedline, was patterned on a room-temperature grown Ti-seeded $\aTa(110)$ film on a tile diced from wafer WA13. The tile was subjected to the optimized cleaning procedure, corresponding to device D5 in Fig.~\ref{fig:fig3}.

Manhattan-style Josephson junctions were fabricated using double-angle evaporation of Al electrodes with intermediate oxidation. A trilayer resist stack of MMA EL8, PMMA 495 A8, and PMMA 950 A4 was first spin-coated. Following exposure of the junction pattern, the resist stack was developed in a 1:3 mixture of MIBK and IPA, then rinsed in two successive IPA baths. Oxygen plasma cleaning was performed to remove residual resist from the developed areas. The junction electrodes were deposited in a Plassys MEB550 electron-beam evaporator, with bottom (top) electrode thickness $30~\nm$ $(100~\nm)$, and intermediate oxidation for $20~\min$ at $1.3~\mBar$. To establish galvanic contact to the capacitor pads, the Al electrodes and Ta base-layer contact regions were ion milled to remove native surface oxides, before depositing an Al bandage layer ($100~\nm)$. Next, the tile was coated with photoresist to protect its surface, diced to size, and the resist lift-off by overnight immersion in NMP at $80\degC$. The finished device was mounted on a custom PCB and wirebonded. To suppress slot-line mode propagation in CPW structures, wirebonds were added across ground-plane gaps on-chip.

This device was measured in Fridge~C (Supplementary Fig.~\ref{fig:supp_fridge_wiring}c) using the cryogenic wiring described in Supplementary Fig.~\ref{fig:supp_fridge_wiring}a. We controlled the device using a Qblox QRM-RF module for readout and a QCM-RFII module for qubit driving, with the two signals combined and delivered through the common feedline. The qubits are $\pi$-pulsed in $500~\ns$ and read out in $4~\us$ using standard dispersive readout~\cite{Blais04}. Standard time-domain routines~\cite{Krantz19} are used to extract the qubit relaxation time $\Tone$ and echo dephasing time $\Techo$. An example single extraction of $\Tone$ of transmon Q6 is shown in Supplementary Fig.~\ref{fig:century}b.  Repeated $\Tone$ measurements of Q6 over 74~h are shown in Supplementary Fig.~\ref{fig:century}c.  Time averages of extracted $\Tone$ and $\Techo$ for all transmons are shown in Supplementary Fig.~\ref{fig:century}d, giving device averages $\Tone=96\pm19~\us$ and $\Techo=148\pm34~\us$.

The contribution of the readout resonator to qubit relaxation (i.e., the Purcell effect) was estimated through measurements of the Lamb shift and the readout resonator linewidth~\cite{Shanto2024}, giving $\TPurcell \sim 240-1200~\us$~\cite{Koch07}. This suggests that Purcell-induced relaxation is not dominant for any qubit, but perhaps not negligible for some.

To estimate the surface-TLS contribution, we simulated the MS participation ratio of the two capacitor variants used, using the method described in Supplementary Sec.~\ref{supsec:PR}, obtaining $\pMS=6.3\times10^{-5}$ and $6.8\times10^{-5}$. With $\tandeff\approx1.64\times10^{-3}$ for the optimal cleaning procedure [Eq.~\ref{eq:tand_eff}], the TLS-limited relaxation time
\begin{equation}
  T_1^\mathrm{TLS}=\frac{1}{\omegaq\,\pMS \tandeff} \approx 270-420~\us
  \label{eq:T1_TLS}
\end{equation}
over the measured qubit frequencies $\omegaq/2\pi=5.3-3.7~\GHz$. The measured $\Tone$ therefore indicates loss beyond the estimated Purcell and surface-TLS contributions, e.g., surface degradation from Josephson-junction processing and loss in the junctions themselves. Distinguishing these contributions will require further systematic study. Nonetheless, with device and time averaged $\Tone$ at the $100~\us$ level,  Ti-seeded, room-temperature-grown $\aTa(110)$ films are shown suitable for full quantum processor fabrication.
\clearpage

\label{sec2}
\end{document}